\documentclass[12pts]{article}
\usepackage{lineno}
\usepackage[dvips]{graphicx}
\usepackage[toc,page]{appendix}
\usepackage{graphicx}
\usepackage{amsmath}
\usepackage[varg]{txfonts}

\usepackage{longtable}

\usepackage[usenames]{color}
\usepackage{xspace}
\usepackage{siunitx}

\usepackage{natbib}
\begin{document}

\title{Modeling the chemical impact and the optical emissions produced by lightning-induced electromagnetic fields in the upper atmosphere: the case of halos and elves triggered by different lightning discharges}

\author{
  F. J. P\'erez-Invern\'on$^{1}$,
  A. Luque$^{1}$,
 F. J. Gordillo-V\'azquez$^{1}$. \\
\textit{$^{1}$Instituto de Astrof\'isica de Andaluc\'ia (IAA),} \\
   \textit{CSIC, PO Box 3004, 18080 Granada, Spain.}
\footnote{Correspondence to: fjpi@iaa.es. 
Article published in Journal of Geophysical Research: Atmospheres. P\'erez ‐ Invern\'on, F. J., Luque, A., and Gordillo ‐ V\'azquez, F. J. (2018). Modeling the chemical impact and the optical emissions produced by lightning‐induced electromagnetic fields in the upper atmosphere: The case of halos and elves triggered by different lightning discharges. Journal of Geophysical Research: Atmospheres, 123, 7615–7641. https://doi.org/10.1029/2017JD028235}
}
\date{}
\maketitle

\begin{abstract}
Halos and elves are Transient Luminous Events (TLEs) produced in the lower ionosphere as a consequence of lightning-driven electromagnetic fields. These events can influence the upper-atmospheric chemistry and produce optical emissions. We have developed different two-dimensional self-consistent models that couple electrodynamical equations with a chemical scheme to simulate halos and elves produced by vertical cloud-to-ground (CG) lightning discharges, Compact Intra-cloud Discharges (CIDs) and Energetic In-cloud Pulses (EIPs). The optical emissions from radiative relaxation of excited states of molecular and atomic nitrogen and oxygen have been calculated. We have upgraded previous local models of halos and elves to calculate for the first time the vibrationally detailed optical spectra of elves triggered by CIDs and EIPs. According to our results, the optical spectra of elves do not depend on the type of parent lightning discharge. Finally, we have quantified the local chemical impact in the upper atmosphere of single halos and elves. In the case of the halo, we follow the cascade of chemical reactions triggered by the lightning-produced electric field during a long-time simulation of up to one second. We obtain a production rate of NO molecules by single halos and elves of 10$^{16}$ and 10$^{14}$ molecules/J, respectively. The results of these local models have been used to estimate the global production of NO by halos and elves in the upper atmopshere at $\sim10^{-7}$ Tg~N/y. This global chemical impact of halos and elves is seven orders of magnitude below the production of NO in the troposphere by lightning discharges.

\end{abstract}

\section{Introduction}
Electromagnetic fields generated by lightning discharges produce Transient Luminous Events (TLEs) in the lower ionosphere, as first proposed by \cite{Wilson1925/PPhSocLon} and later detected by \cite{Franz1990/Sci}. The local chemical impact and optical signature of these events have been investigated by many authors \citep{Sentman2008/JGRD/1, Gordillo-Vazquez2008/JPhD, Gordillo-Vazquez2010/JGRA, Gordillo-Vazquez2011/JGR, Gordillo-Vazquez2012/JGR, Parra-Rojas/JGR, Parra-Rojas/JGR2015, Kuo2007/JGRA, Winkler2015/JASTP}. Some recent estimations suggest a future enhancement in the lightning activity as a consequence of the global temperature increase \citep{Romps2014/SCI}. Therefore, the study of the chemical influence of electric discharge processes in the atmosphere emerges as an important field in the characterization of the future atmosphere. Models and observations are needed to quantify the impact of these events. 

Based on the physical production mechanism, TLEs can be categorized into elves, halos, sprites, blue jets, or gigantic jets \citep{Pasko2012/SSR}. Halos and sprites are a direct consequence of the quasi-electrostatic field created by lightning discharges \citep{Pasko2012/SSR}, while elves (Emission of Light and Very low frequency perturbations due to Electromagnetic pulse Sources), discovered by \cite{Fukunishi1996/GeoRL}, are very fast TLEs produced by the interaction between lower ionospheric electrons and lightning-emitted ElectroMagnetic Pulses (EMPs).

In this paper, we will focus on halos and elves. Halos are glow discharges usually produced at altitudes between 75 km and 80 km when the reduced electric field created by lightning discharges reaches the breakdown value of $\sim$120 Td \citep{Barrington-Leigh2001/JGR,Wescott2001/JGR/1, Bering2002/AdSpR,Moudry2003/JASTP, Luque2011/NatGe}. When the electric field becomes larger than the breakdown value, the electron-driven ionization rate of air molecules becomes larger than the attachment rate, causing a high increase in the electron density. Halos are usually seen accompanying a sprite, another kind of TLE formed by a complex structure of streamers. Some observations of single halos have been reported \citep{Marshall2006/JGRD, Kuo2013/JGR}. The optical emissions from halos usually last between one and three milliseconds. This kind of TLE is usually associated with negative or positive cloud-to-ground (CG) lightning discharges \citep{Bering2002/AdSpR,Bering2004/AdSpR,Bering2004/GeoRL,Frey2007/GeoRL}, since the return stroke stage of the CG lightning can transfer more total charge than intra-cloud (IC) or cloud-to-cloud (CC) lightning discharges \citep{Rakov2003/ligh.book,Maggio2009/JGR}.

Elves are very fast optical emissions (less than 1 ms) as a consequence of the heating of electron produced by lightning-generated EMPs \citep{Inan1991/GRL, Inan1997/GeoRL,Moudry2003/JASTP}. This type of TLE is usually observed in a thin layer of the upper atmosphere usually located at 88 km of altitude \citep{van_der_Velde2016/GRL}, with a radius of more than 200 km. Lightning detection networks associate most of observed elves with cloud-to-ground (CG) lightning discharges, whose electromagnetic emission pattern produces toroidal light emissions in the lower ionosphere \citep{Moudry2003/JASTP, van_der_Velde2016/GRL}. 

The first observation of elves from spacecraft was accomplished by the Space Shuttle \citep{Boeck1992/GRL}. Other space missions, such as ISUAL \citep{Chang2010/JGRA} and JEM-GLIMS \citep{Adachi2016/JASTP}, have investigated the optical signature of elves, mainly recording photons from the first and second positive systems of the molecular nitrogen (FP system of N$_2$ and the SP system of N$_2$),  first negative system of the molecular nitrogen ion (N$_2$$^+$-1N), Meinel band of the molecular nitrogen ion (Meinel N$_2$$^+$) and Lyman-Birge-Hopfield (LBH) band of the molecular neutral nitrogen. In addition, ground-based observations have provided accurate information about the altitude and structure of elves \citep{van_der_Velde2016/GRL}. Moreover, some authors have previously investigated elves by modelling the impact of EMPs in the lower ionosphere \citep{Inan1991/GRL, Taranenko1993/GRL, Kuo2007/JGRA, Marshall2010/JGRA/2, Marshall2012/JGR}. 

Recent investigations have also related elves with Compact Intra-cloud Discharges (CIDs) \citep{Marshall2015/GRL} and Energetic In-cloud Pulses (EIPs) \citep{Liu2017/JGRAinpress}. These kind of events are very fast and powerful discharges, first reported by \cite{Levine1980/JGR}, producing high electric field enhancements recorded some tens of kilometers away. Although the mechanism behind these events is not fully understood, some models explain the electromagnetic signals related to these types of lightning discharges by coupling of relativistic runaway electron avalanche (RREA) and extensive atmospheric shower of cosmic rays with strong localized electric fields \citep{Gurevich2001/PhyU, Gurevich2004/PLA, Watson2007/GRL}. The total charge moment change (CMC) produced by CIDs and EIPs is too low (below 3 C km \citep{Karunarathne2016/JGR}) to trigger halos or sprites. However, the pulses originated by both CIDs and EIPs can produce elve doublets after reflection with the ground surface, as previously studied by \cite{Marshall2015/GRL} and \cite{Liu2017/JGRAinpress}. \cite{Gaopeng2010/GRL} proposed the study of these events to confirm or discard their relation with Terrestrial Gamma-ray Flashes (TGFs).

The detailed study of the chemical impact of halos and elves and their relation with the parent lightning discharge is still an open research topic. In the near future, two space-based missions will be devoted to the study of TLEs. The Atmosphere-Space Interactions Monitor (ASIM) \citep{Neubert2006/ILWS} of the European Space Agency (ESA) was successfully launched on April 2, 2018 to observe TLEs from the International Space Station (ISS). In addition, the Tool for the Analysis of RAdiations from lightNIngs and Sprites (TARANIS) \citep{Blanc2007/AdSpR} mission of the Centre National d'\'Etudes Spatiales (CNES) will be launched in 2019. Both ASIM and TARANIS are equipped with TGF detectors and with high-temporal resolution photometers capable of recording optical signals fromTLEs.

 In this paper, we contribute to the knowledge of TLEs by modeling the inception and evolution of halos and elves using two different electrodynamical models that share the same set of kinetic reactions. On the one hand, we develop a self-consistent 2D model of lightning-produced quasi-electrostatic fields coupled with the kinetic scheme to simulate the local chemical signature produced by halos triggered by vertical CG lightning discharges. This halo model is based on the same scheme proposed by \cite{Pasko1995/GeoRL} and continued by other authors \citep{Luque2009/NatGe,Neubert2011/JGRA,Liu2015/NatCo, Qin2014/NatCo,PerezInvernon2016/GRL, PerezInvernon2016/JGR, Kabirzadeh2015/JGR, Kabirzadeh2017/JGR}. We have contributed to the upgrade of these models by adding a more detailed chemical scheme and extending the chemical simulations up to times of one second after the onset of the halo. On the other hand, we have developed a self-consistent Finite-Difference Time Domain (FDTD) model of electromagnetic wave propagation that has also been coupled with the kinetic scheme to simulate elves produced by vertical CG lightning discharges, CIDs and EIPs. This FDTD model is based on the one proposed by  \citep{Inan1991/GRL} and later on upgraded by other authors \citep{Taranenko1993/GRL, Kuo2007/JGRA, Marshall2010/JGRA/2, Marshall2012/JGR}. \cite{Marshall2015/GRL} and \cite{Liu2017/JGRAinpress} investigated the elves triggered by impulsive in-cloud dicharges, such as CIDs and EIPs, using a basic chemical scheme. In this work, we use a similar FDTD scheme to simulate the electrodynamic of elves coupled with a detailed set of chemical reactions. This approach has allowed us to compare the spectra of different types of lightning generated very impulsive TLE.

In section~\ref{sec:models} we describe the developed models, their implementation for different source currents and the method to compute the synthetic spectra of the predicted TLEs. In section~\ref{results} we present and discuss the results of our calculations. We quantify the chemical influence of halos in the atmosphere and compare the calculated spectra of each simulated TLE. Finally, we conclude in section~\ref{conclusions} summing up the most important implications of the results obtained.

\section{Models} \label{sec:models}

In section~\ref{kineticmodel} we describe the kinetic scheme that determines the local chemical impact and optical emissions due to halos and elves. The detailed electrodynamical model developed for each case is detailed in sections~\ref{sec:modelhalos} and~\ref{sec:FDTD}

\subsection{Kinetic scheme} \label{kineticmodel}

The kinetic scheme determines the interaction between particles, as well as electron mobility and diffusion in the presence of a reduced electric field. We use a kinetic scheme with 136 species interacting through 1076 chemical reactions. The complete set of chemical reactions is detailed in the supporting information \citep{Gordillo-Vazquez2008/JPhD, Sentman2008/JGRD/1, Parra-Rojas/JGR, Parra-Rojas/JGR2015, Gordillo-Vazquez2010/JGRA, Phelps1991/JPCRD, Cartwright1977/PhRvA, Borst1970/PhRvA, Simek2002/JPhD/1, Lawton1978/JCP, Pagnon1995/JPD, Erdman1987/JCP, Skalny1996/CPL,Morgan2000/AAMOP, Laher1990/JPCRD, Hayashi1985/proc, phelps1969, PhelpsCO2,Pancheshnyi2013/JAP, Kossyi1992/PSST, Peverall2001/JChPh, Whitaker1981/PRA, Starikovskaia2001/CTM, Castillo2004b/PSST, Gudmundsson2001/JPD, Rodriguez1991/JAP, Gordiets1995/ITPS, Kamaratos2006/CP, yaron1976/CPL, slanger2003/CR, VallanceJones1974/book/1, viggiano2006/JPCA, Guerra1999/PSST, Zinn1990/JGR, Herron2001/PCPP, Brasseur2nd/book, Makhlouf1995/JGR, Turnbull1991/GeoRL, cacciatore2005/JCP, kurnosov2007/JPCA, Capitelli/Book, linstrom2015nist, Simek2003/PSST, Thoman1992/JChPh/c, biondi1971defense, kazil2003university, albritton1978ion, Piper1985/JChPh, Lepoutre1977/CPL, Dagdigian1988/CPL, Morrill1996/JGR, Calo1971/JChPh/1, Piper1988/JChPh, Piper1992/JChPh/c, Piper1989/JChPh, kam1991/PRA,smirnov1982negative,BATES1988869,Krupeine1972/JPCRD,radzig2012reference,Gilmore1992/JPCRD}.
 The main reaction types in terms of the chemical and optical impact produced by TLEs are:

\begin{enumerate} 
\item Electron impact excitation of neutral species given by

\begin{linenomath*}
\begin{equation}
e + A \rightarrow e + A^{*},
\end{equation}
\end{linenomath*}

and ionization given by

\begin{linenomath*}
\begin{equation}
e + A \rightarrow 2e + A^{+},
\end{equation}
\end{linenomath*}
where the species A can be N$_2$, O$_2$, N, O, NO, N$_2$O, O$_3$ or CO$_2$. The species A$^{*}$ can be electronically and/or vibrationally excited in the case of N$_2$, while only electron excitation is considered for the rest of species. A$^{+}$ stands for the positive ion of the molecule or atom A. The cross sections used to calculate these reaction rates are detailed in the supporting information.

\item Electron attachment processes, among which the most important is the electron driven dissociative attachment of O$_2$ molecules, 
\begin{linenomath*}
\begin{equation}
e + O_2 \rightarrow O + O^{-}.
\end{equation}
\end{linenomath*}
This attachment reaction rate dominates electron-ionization processes of N$_2$ and O$_2$ for reduced electric fields below the breakdown value. When the breakdown field is reached, ionization becomes larger than attachment, triggering a cascade of reactions that, ultimately, lead to the observation of TLEs.

\item Electron detachment processes. As previously studied by \cite{Luque2011/NatGe, Marshall2012/JGR}, some detachment processes like associative detachment of O$^-$ by N$_2$ can dominate the production of electrons at electric fields below the breakdown value. We consider the electron detachment from negative ions by N$_2$ and O$_2$ described by \cite{Pancheshnyi2013/JAP}, that are electric-field-dependent. Furthermore, we include also electron detachment from O$^-$ interacting with other important species, such as CO and NO \citep{biondi1971defense, Kossyi1992/PSST}. However, according to \cite{Moruzzi1968/TJCP}, the electric field dependence of O$^-$ + CO and O$^-$ + NO detachment rates is negligible.

\item Electron-ion and ion-ion recombination processes, contributing to remove charge carriers \citep{Gordillo-Vazquez2008/JPhD,Sentman2008/JGRD/1, Parra-Rojas/JGR, Parra-Rojas/JGR2015}.

\item Vibrational redistribution, energy pooling, Vibrational-Translational (VT) and Vibrational-Vibrational (VV) processes involving electronically and vibrationally excited molecules of N$_2$ \citep{Gordillo-Vazquez2010/JGRA}.

\item Chemical reactions involving excited state species \citep{Gordillo-Vazquez2008/JPhD,Sentman2008/JGRD/1, Parra-Rojas/JGR, Parra-Rojas/JGR2015}.

\item Positive and negative ion chemistry, in addition to ground state chemistry. These reactions contribute to the enhancement of some neutral species \citep{Gordillo-Vazquez2008/JPhD,Sentman2008/JGRD/1, Parra-Rojas/JGR, Parra-Rojas/JGR2015} .

\item Odd hydrogen and odd nitrogen reactions \citep{Sentman2008/JGRD/1}.

\item Radiative decay, electronic quenching and vibrational quenching. These three processes compete to de-excitate molecules and atoms. Most important quenching reactions include N$_2$ and O$_2$. The most of the radiative decay constants and electronic and vibrational quenching rates used in this work are taken from \citep{Gordillo-Vazquez2010/JGRA}. However, to the best of our knowledge, the electronic quenching rate of N$_2$(E$^3$$\Sigma ^+ _g$) by N$_2$ or O$_2$ is not described in the literature. This process can be important to describe the Vibrational Distribution Function (VDF) of the electro-vibrationally excited molecule N$_2$(C$^3$$\Pi _u$ , v). As an approximation of the electronic quenching rate of N$_2$(E$^3$$\Sigma ^+ _g$) we set its value equal to the quenching rate of the molecule N$_2$(C$^3$$\Pi _u$ , v = 0).

\end{enumerate}

Mobility, diffusion and some reaction coefficients contributing to species production and loss can also depend on the reduced electric field. These dependences are obtained by solving off-line the steady-state Boltzmann equation for the gas mixture (humid air) using the packaage BOLSIG+ \citep{Hagelaar2005/PSST}.

We calculate the temporal evolution of the concentration of emitting atoms and molecules computing the total emitted photons per second according to the radiative processes. This kinetic scheme allows us to predict the existence or absence of halo optical emissions in some important spectral bands and lines, such as the nitrogen first  (550 nm - 1200 nm) and second positive (250 nm - 450 nm) systems, nitrogen first negative line between the zero vibrational leves v$^\prime$ and v$^{\prime\prime}$ (391.4 nm), nitrogen Lyman-Birge-Hopfield (LBH) band (110 nm - 200 nm), molecular oxygen atmospheric (538 nm - 1580 nm), Noxon (1908 nm) and Herzberg I (243 nm - 488 nm) systems, atomic oxygen green (557 nm), red (630 nm), near infrared lines (777 nm and 844 nm), and other atomic and molecular emissions. We pay special attention to the vibrational chemistry of excited species such as N$_2$(B$^3$$\Pi _g$ , v = 0,...,6), N$_2$(C$^3$$\Pi _u$ , v = 0,...,4) and N$_2$(a$^1$$\Pi _g$ , v = 0,...,15). This detailed description of the vibrational levels of the mentioned excited species allows us to estimate their VDFs. The proposed kinetic scheme is also useful to predict the optical emission spectra of halos and elves corresponding to the first and second positive systems of molecular nitrogen, as well as to the LBH band. We collect the emission bands in the last column of table~\ref{tab:species}.

Finally, we use the software QTPlaskin, developed by \cite{QTplaskin}, to analyze the main processes that contribute to create or destroy each species as a function of time.

\subsubsection{Optical emissions}

Let $N_i$($\vec{r}$) be the spatial density distribution of one of the calculated species N$_2$(B$^3$$\Pi _g$ , v = 0,...,6), N$_2$(C$^3$$\Pi _u$ , v = 0,...,4), N$_2$(a$^1$ $\Pi _g$ , v = 0,...,15), NO(A$^2\Sigma^+$), O$_2$(A$^3\Pi_u^+$), O$_2$(b$^1\Pi_g^+$), O$_2$(a$^1\Delta_g$), O($^1$S), O($^1$D), O($^3$P) or O($^5$P), and A$_i$ the radiative decay constant of de-excitation and emission of photons at a given wavelength. The temporal evolution of the total number of emitted photons per second from each species can be calculated integrating in volume as

\begin{linenomath*}
\begin{equation}
I = \int A_i N_i(\vec{r}) dV. \label{emissions}
\end{equation}
\end{linenomath*}

The knowledge of these quantities together with the energy difference between levels allow us to build synthetic spectra. As a consequence of the low air density above TLEs, the observed spectra from spacecraft would be very similar to the spectra at the emitting source. However, if the spectrograph is located near the ground level, as in the case of the instrument "\textit{GRAnada Sprite Spectrograph and Polarimeter}" (GRASSP) \citep{passas2014/IEEE, Passas2016/APO, Gordillo-Vazquez2018/JGR}, the effect of the atmospheric attenuation on each spectral transition has to be included in the calculation of the emitted spectra in order to compare with observed spectra.

We calculate the optical transmittance of the atmosphere between an emitting TLE and an observer located at a horizontal distance of 350~km using the software MODTRAN~5 \citep{berk2005modtran}. As can be seen in figure~2 of \cite{Gordillo-Vazquez2012/JGR}, the air transmittance dependence on the light frequency is irregular. We use a tool code previously developed in the study of \cite{Parra-Rojas2013/JGR} to obtain the rovibronic bands of the FP system of N$_2$ system. This program is based on the calculation of the decay constant of each rovibrational level following the method described in \cite{Kovacs1969/book} for triplet transitions. Afterwards, the obtained transmittance for different altitudes can be applied to the emitted spectra to derive the predicted observed optical signature of halos and elves from ground-based spectrographs like GRASSP \citep{Passas2016/APO}.

\subsection{Model of halo production} \label{sec:modelhalos}

We model the impact of lightning-induced quasi-electrostatic fields in the lower ionosphere of the Earth using a cylindrically symmetrical scheme similar to the one used in previous models by e.g. \cite{Luque2009/NatGe,Neubert2011/JGRA,Liu2015/NatCo, Qin2014/NatCo,PerezInvernon2016/GRL} and \cite{PerezInvernon2016/JGR}. The time evolution of the electric field is coupled with the transport of charged particles and with an extended set of chemical reactions. 

Strokes produced by CG lightning discharges lower electric charge to the ground, accumulating the same amount of opposite sign charge at cloud altitudes. We model this charge accumulation as a sphere of radius 0.5 km at an altitude $h=\SI{7}{km}$ \citep{Maggio2009/JGR}. The time evolution of the total accumulated charge is given by a bi-exponential function

\begin{linenomath*}
\begin{equation}
\frac{dQ(t)}{dt} = I(t) = \frac{Q_{\text{max}}}{\tau_1 - \tau_2} \left(\exp(-t/\tau_1)-\exp(-t/\tau_2)\right), \label{Q}
\end{equation}
\end{linenomath*}
where $Q_{\text{max}}$ is the opposite sign total charge lowered to the ground and where $\tau_1=\SI{1}{ms}$ and $\tau_2=\SI{0.1}{ms}$, respectively, the total discharge time and the rise time of the discharge current. The charge moment change (CMC), given by the product $hQ_{\text{max}}$, determines both the electric field imposed on the ionosphere and its strength.

Halos are a direct consequence of the lightning-induced quasi-electrostatic field. The shape of the optical emissions produced by a halo corresponds to the spatial distribution of the quasi-electrostatic field created right above a dipolar lightning discharge. In addition, the characteristic time of halos (several milliseconds) also agrees with the characteristic time of the quasi-electrostatic field. For this reason, we neglect the effect of the radiation field. The quasi-electrostatic field created by the charge resulting from equation (\ref{Q}) is calculated using FISHPACK \citep{Fishpack}, a 2-D~Poisson solver in cylindrical coordinates that solves the equation

\begin{linenomath*}
\begin{equation}
\nabla^2 \phi = - \frac{\rho}{\epsilon_0}, \label{Poisson}
\end{equation}
\end{linenomath*}

where $\phi $ is the electric potential, $\epsilon_0$ is the permitivity of vacuum and $\rho$ is the charge density located in the integration domain. This quantity includes the lightning-accumulated charge in the troposhere as well as the induced charge in the mesosphere and lower ionosphere.

The resultant electric field induces electron and ion transport in the mesosphere and lower ionosphere, separating charges of opposite sign. The transport of the charged species $i$ is determined by the advection-diffusion flux $\mathbf{\Phi}_i$, whose components in the vertical and radial directions ($z$ and $r$) are given, in the case of electrons, by

\begin{subequations}
\label{transportelectrons}
\begin{linenomath*}
\begin{align}
\Phi_{e, z} = -D_e \frac{\partial N_{e}}{\partial z} - \mu_{e} E_z N_{e} \\
\Phi_{e, r} = -D_e \frac{\partial N_{e}}{\partial r} - \mu_{e} E_r N_{e}, 
\end{align}
\end{linenomath*}
\end{subequations}

while in the case of ions we can neglect diffusion and write these equations as

\begin{subequations}
\label{transportions}
\begin{linenomath*}
\begin{align}
\Phi_{i, z} = - v_{i,z} N_i ,\\
\Phi_{i, r} = - v_{i,r} N_i, 
\end{align}
\end{linenomath*}
\end{subequations}

where $D_e$, $\mu_{e}$ and $N_{e}$ are, respectively, the electron diffusion coefficient, mobility and density, while $v_i$ and $N_{i}$ are the ion velocities and densities. The electron diffusion coefficient and mobility depend on the gas composition and the electric field, and will be described in the next section. In the case of ions, the velocity of an ion with mass $m_i$ under the influence of high reduced electric fields is given by \citep{Fahr1967/ZPHN, Pancheshnyi2013/JAP}

\begin{linenomath*}
\begin{equation}
v_i = \sqrt{\frac{2 \theta_z}{\pi m_i}}, \label{ionvelocity}
\end{equation}
\end{linenomath*}

where $\theta_z$, the kinetic energy of the ion in the presence of an electric field $E$, is given by \cite{Fahr1967/ZPHN, Pancheshnyi2013/JAP} as

\begin{linenomath*}
\begin{equation}
\theta_z = q_e E \lambda, 
\end{equation}
\end{linenomath*}

where $q_e$ is the negative elementary electric charge and $\lambda$ is the mean free path of particles in the gas between successive collisions, calculated as a function of the air density $N$ \citep{chapman1970mathematical}

\begin{linenomath*}
\begin{equation}
\lambda = \left(\sqrt{2} \pi d^2 N\right)^{-1}. 
\end{equation}
\end{linenomath*}

where $d$ is the average diameter of molecules in air.

In the case of halos, where the quasi-electrostatic field remains high for several milliseconds, numerical oscilations can appear as a consequence of large density gradients. For this reason, equations (\ref{transportelectrons}) and (\ref{transportions}) are solved using a Koren limiter function \citep{Montijn2006/JCoPh} to obtain charged species fluxes. Then, we are able to write the continuity equation of the species $i$ including kinetics and transport as

\begin{linenomath*}
\begin{equation}
\frac{\partial N_{e,i}}{\partial t} + \nabla \cdot \mathbf{\Phi}_{e,i} = P_{e,i} - L_{e,i},
\label{continuity}
\end{equation}
\end{linenomath*}

where $P_{e,i}$ and $L_{e,i}$ are the production and loss rates of electrons or species $i$, determined by the kinetic scheme. We solve this equation using an explicit Runge-Kutta method of order 5 with step size control based on the Dormand and Prince algorithm \citep{DormandPrince1980/JAM}. We solve this equation for altitudes above 50~km, where the plasma density is important.

Equations (\ref{Poisson})-(\ref{continuity}) are discretized in a cylindrical grid with a spatial resolution given by $\Delta z$~=~100~m and $\Delta r$~=~500~m. We use an adaptive time step $\Delta t$ $<$ 3 $\times$ 10$^{-6}$ s. For the Poisson equation (\ref{Poisson}), we use boundary conditions of the Neumann type at $r$~=~300~km, undetermined in the axis of symmetry and of the Dirichlet type in the upper (ionosphere) and lower (ground) boundaries assuming that both of them behave as perfect electric conductors.
In the case of the transport equations (\ref{transportelectrons}) and (\ref{transportions}) we use Neumann type boundary conditions.

The quasi-electrostatic field produced by the discharge evolves influenced by the total transferred charge (\ref{Q}) roughly during the first millisecond. After that time, the lightning discharge ends and no more charge is deposited on the clouds. Therefore, the main change of the quasi-electrostatic field in the lower ionosphere is determined by the field-induced electron currents, with time scales approximately given by the dielectric relaxation time at each altitude level. Therefore, the time derivative of the quasi-electrostatic field will decrease progressively as the field is screened at higher altitudes. We take advantage of this fact and implement a method to progressively decrease the computational cost of our simulation. We include a parameter $p$ in the code that calculates the maximum time derivative of the reduced electric field at a given time t$^\prime$ by computing the difference between the field at that time and at the previous time step (t$^\prime$-$\Delta$t):

\begin{linenomath*}
\begin{equation}
p\left(t^\prime\right) = \max\left( \frac{ \partial\left(\frac{ E\left(r,z,t^\prime\right)}{N(r,z)}\right) }{\partial t^\prime}  \right)  = \max\left( \frac{ \frac{E\left(r,z,t^\prime\right) - E\left(r,z,t^\prime-\Delta t \right)}{N(r,z)} }{\Delta t}  \right). \label{derivative}
\end{equation}
\end{linenomath*}

This parameter gives us the maximum absolute variation of the reduced electric field per second. Using $p$ we can then estimate the total number of time steps $n$ that are necessary to see a maximum absolute change of 0.1~Td as

\begin{linenomath*}
\begin{equation}
 n \Delta t = \frac{0.1}{p}. 
\end{equation}
\end{linenomath*}

If $n$ is greater than one, we deactivate the Poisson equation solver for that number of subsequent time steps, keeping the electric field and its effects constant. This allows us to exclusively solve the equations related to the chemistry and the charged particles transport assuming a maximum relative error in the electric field less than 0.1~Td.

\subsection{Model of elve production} \label{sec:FDTD}

As mentioned in the introduction, we have also developed a two-dimensional Finite-Difference Time Domain (FDTD) model of electromagnetic wave propagation coupled with a chemical scheme. This FDTD model solves the Maxwell equations in a  2-D grid domain \citep{Inan1991/GRL, Taranenko1993/GRL, Kuo2007/JGRA, Marshall2010/JGRA/2, Inan2011/Book, Luque2014/JGRA, Marshall2015/GRL, Liu2017/JGRAinpress} to obtain the electric and magnetic field vectors $\mathbf{E}$ and $\mathbf{H}$ produced by lightning discharges (CG, CID or EIP). In particular, the developed FDTD model is implemented in a cylindrical 2-D grid domain.The model uses a modified Ohm's equation to calculate the current density induced by the electric fields in the upper atmosphere \citep{Lee1999/IEE, Luque2014/JGRA}. We couple the scheme of electromagnetic wave propagation with a set of continuity reactions, updating component densities $n_i$ at each time step. The complete set of equations is given by

\begin{linenomath*}
\begin{equation}
\bigtriangledown \times \mathbf{E} = -\mu_0 \frac{\partial \mathbf{H}}{\partial t}, \label{maxwell1}
\end{equation}
\begin{equation}
\bigtriangledown \times \mathbf{H} = \epsilon_0 \frac{\partial \mathbf{E}}{\partial t} + \mathbf{J}, \label{maxwell2}
\end{equation}
\begin{equation}
\frac{d \mathbf{J}}{d t}+ \nu \mathbf{J} = \epsilon_0 \omega_{p}^{2} (\mathbf{r},t) \mathbf{E} + \mathbf{\omega}_{b} (\mathbf{r},t) \times \mathbf{J} , \label{lang}
\end{equation}
\begin{equation}
\frac{\partial N_i}{\partial t} = G_i - L_i . \label{cont}
\end{equation}
\end{linenomath*}

We solve Maxwell equations (\ref{maxwell1}) and (\ref{maxwell2}), where $\epsilon_0$ and $\mu_0$ are the permitivity and permeability of free space, using a two-dimensional FDTD model in a cylindrical 2-D grid domain using the Yee algorithm \citep{Yee1966/IEEE} with a space step $\Delta d$ shorter than the minimum characteristic wavelength of the source electric current in each case and with a time step shorter than $\Delta d / \sqrt{3}c$ \citep{Inan2011/Book}. The term $\mathbf{J}$ contains current densities, i.e., the lightning channel current density and the electron current density induced by electric fields in the lower ionosphere. We impose absorbing boundary conditions using convolutional perfectly matched layers \citep{Inan2011/Book}. We define the ground as a perfect conductor.

The modified Ohm's equation~(\ref{lang}) is solved using the same two-dimensional FDTD model as the Maxwell equations, as proposed by \cite{Lee1999/IEE}. Equation~(\ref{lang}) is only solved at altitudes where the electron density becomes important. At these altitudes near the ionosphere, the electron conductivity is orders of magnitude higher than the ion conductivity, hence we neglect the ion current density contribution. We use the same notation as \cite{Lee1999/IEE} and \cite{Luque2014/JGRA}, where $\nu = e/\mu m_e$ is the effective collision frequency between electrons and neutrals, dependent on electron charge magnitude $e$ and mass $m_e$, and on electron mobility $\mu$. The term $\omega_p = (e^2n_e / m_e \epsilon_0)^{1/2}$ corresponds to the plasma frequency for electrons and depends on the electron density $n_e$. Finally, $\mathbf{\omega}_b = e\mathbf{B}_0 / m_e$ is the electron gyro frequency, where $\mathbf{B}_0$ is the background magnetic field, considered zero in the two-dimensional approximation.

\cite{Marshall2008/AGUFMAE13, Marshall2010/JGRA/2, Marshall2014/JGR} investigated the effect of the background magnetic field in the wave attenuation. The angle between the direction of wave progagation of lightning-radiated waves and the magnetic field vector, as shown by \cite{Marshall2008/AGUFMAE13, Marshall2010/JGRA/2, Marshall2014/JGR}. Considering that lightning-radiated waves are upward, the angle between their direction of propagation and the background magnetic field is determined by the geomagnetical latitude.  As discussed by \cite{Marshall2008/AGUFMAE13, Marshall2010/JGRA/2, Marshall2014/JGR}, the effect of the geomagnetic field can be neglected for angles ranging between 0$^\circ$ and 45$^\circ$, but can be important for higher geomagnetic latitudes. As the occurrence of lightning is mostly gathered at tropical latitudes \citep{Christian2003/JGR}, we neglect the effect of the geomagnetic field. However, it is worth emphasizing that the propagation of the pulses and the subsequent elves would be different at higher geomagnetic latitudes \citep{Marshall2008/AGUFMAE13, Marshall2010/JGRA/2, Marshall2014/JGR}.

%The lightning-radiated pulses can reach higher altitudes when the wave propagation is in the same direction than the background magnetic field vector, while it would suffer stronger attenuation if the propagation is perpendicular to the magnetic field. As discussed by \cite{Marshall2008/AGUFMAE13, Marshall2010/JGRA/2, Marshall2014/JGR}, the effect of the magnetic field can be neglected if the angle between the geomagnetic field and the 
%As discussed by \cite{Luque2014/JGRA}, we can neglect the effect of the background magnetic field $\mathbf{B}_0$ in equation (\ref{lang}) at altitudes where elves are produced.

Equation (\ref{cont}) describes the evolution of each component's density as a function of its gains $G_i$ and losses $L_i$. This equation will be particularized for each component and will be coupled with equations (\ref{maxwell1}), (\ref{maxwell2}) and (\ref{lang}) as a consequence of the electric field dependence of some reaction rates. We solve this equation using a forward Euler method choosing a time step smaller than the dielectric relaxation time, the fastest chemical reaction characteristic time and $\Delta d / \sqrt{3}c$.

Regarding the CG lightning discharge, we assume that the return stroke current follows a bi-exponential function \citep{Rakov2003/ligh.book} of the form

\begin{linenomath*}
\begin{equation}
I(t) = I_0 \left(\exp(-t/\tau_1) - \exp(-t/\tau_2)\right), \label{current}
\end{equation}
\end{linenomath*}
where $\tau_2$ is the rise time of the current wave, and $\tau_1$ is the total duration of the stroke. According to \citep{Rakov2003/ligh.book}, $\tau_2$ and $\tau_1$ have characteristic values of tens of microseconds and hundreds of microseconds, respectively. In this work, we set a risetime of \SI{5}{\micro\second} and a total duration of the stroke of \SI{50}{\micro\second} \citep{heidler1999calculation}.

We set a 7~km long channel propagation velocity at 0.75 $\times$ $c$, where $c$ is the speed of light. The vertical source current density vector $\mathbf{J}(t)$ is obtained dividing the lightning current $I(t)$ by the lateral section of a cell. We have considered CG lightning discharges with current peaks of 154~kA, 220~kA, 275~kA and 440~kA \citep{Rakov2003/ligh.book}, respectively. According to \cite{Barrington-Leigh1999/GeoRL/1}, the lightning peak current threshold for the production of elves is about 60~kA, while new results suggest that this threshold is 88~kA \citep{Blaes2016/JGR}. The detection threshold for elves of the ISUAL instrument is estimated to be around 80~kA \citep{Kuo2007/JGRA, Chern2014/AA}.

In the case of elves triggered by CIDs located at 18~km of altitude, we set as source of EMP the lightning currents modeled by \cite{Watson2007/GRL}. We use the Modified Transmission Line Exponential Increasing (MTLEI) model proposed by \cite{Watson2007/GRL} for downward positive discharges, with a peak current of $\sim$400~kA \citep{Cummer2014/GRL, Lyu2015/GRL, Liu2017/JGRAinpress}.  
The EIP negative current source located at 13~km of altitude is taken from \cite{Liu2017/JGRAinpress}, who simulated a EIP-driven elve produced by a current with a peak of $\sim$542~kA. As \cite{Liu2017/JGRAinpress} claims, this current could produce TGFs that could be detected by Fermi \citep{Briggs2010/JGR, Cummer2014/GRL, Lyu2015/GRL}.

After defining the atmosphere composition and the current source, we solve the system of equations (\ref{maxwell1})-(\ref{cont}) in a two-dimensional mesh where $r$ corresponds to horizontal distances from the lightning discharge and $z$ corresponds to altitude. Horizontal distances are between 0~km and 550~km with a step of 0.5~km in the case of elves triggered by CG lightning discharges and between 0~km and 250~km with a step of 0.1~km in the case of elves triggered by CID and EIPs. The altitude domain is between 0~km and 97~km, with a vertical step of 0.1~km. We include 20-cell-wide absorbing boundaries. Equations (\ref{lang}) and (\ref{cont}) are exclusively solved in the region where electron density is important, that is, above 50~km of altitude. Regarding the time step, we set it to 10~ns, ensuring that the constrains detailed above are satisfied.

We have developed this method in several Fortran subroutines, compiling them to create Python modules. The code is parallelized with a shared-memory approach based on OpenMP. We run each parallelized simulation describing 1~ms of the evolution of the elve during a time of about one day using 12~CPUs.

\subsection{Initial conditions}

\begin{figure}
\includegraphics[width=1\columnwidth]{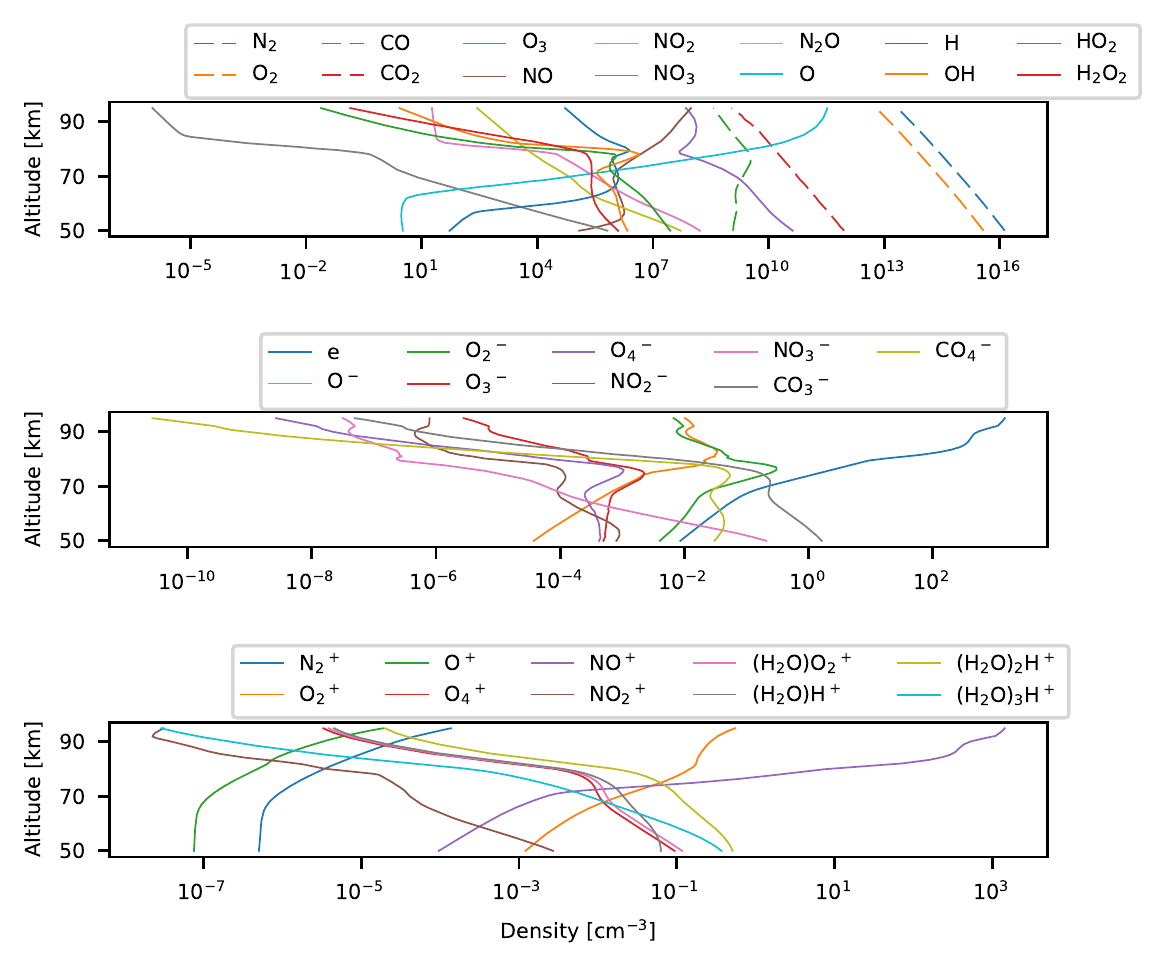}
\caption{\label{fig:equilibrium}
Most important species at equilibrium conditions. Dashed lines correspond to most abundant species. The H$_2$O density is set to 1.6$\times$10$^9$~cm$^{-3}$.
}
\end{figure}

As initial conditions, we use the air density profile and composition at nighttime conditions in November from the Whole Atmosphere Community Climate Model (WACCM) \citep{Marsh2013/JC} for a latitude of 38$^{\circ}$ and a longitude of 0$^{\circ}$. We also use the electron density proposed by \cite{Hu2007/JGRD}. Then we relax the system for 6.5 s under the presence of cosmic ray ionization \citep{Thomas1974/RaSc} solving the continuity equation of each particle to obtain the equilibrium profiles associated to our model. The concentrations of N$_2$, O$_2$, CO, CO$_2$ and H$_2$O are assumed to be constant. Figure \ref{fig:equilibrium} shows the most important species at equilibrium conditions.

\section{Results and Discussion} \label{results}

\subsection{Halo model results}

We apply the model described in section~\ref{sec:modelhalos} to simulate the inception of halos and to quantify both their local chemical impact in the upper atmosphere and their optical signature.

Ionization, attachment and other reaction rates have a non-linear dependence with the reduced electric field. Due to this fact, the upper atmospheric chemical influence and optical emissions driven by lightning depend strongly on the Charge-Moment-Change (CMC) associated with the discharge. We investigate the effect of three different CG lightning discharges with CMCs of 140~C km, 350~C km and 560~C km \citep{MarshallTC2001/JGR, Rakov2003/ligh.book, Cummer2004/GeoRL, Maggio2009/JGR}. We also explore the ``long" time (1~second) halo mesospheric chemistry for the last case.

\subsubsection{Local chemical influence of halos in the mesosphere}
\label{sec:halochemical}
\begin{figure}
\includegraphics[width=1\columnwidth]{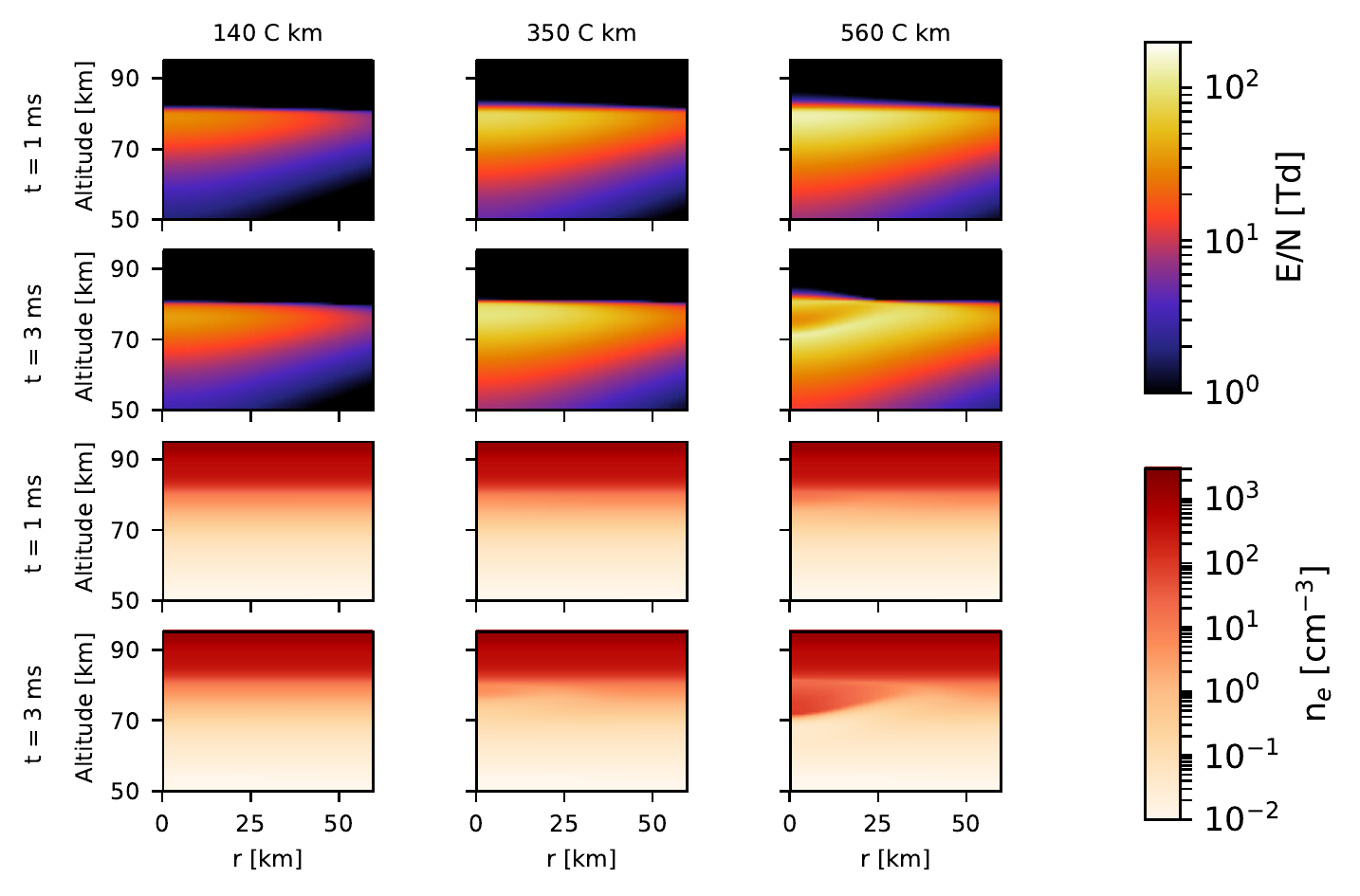}
\caption{\label{fig:Halos_Ered}
Reduced electric field and electron density in the upper atmosphere produced by three different vertical CG lightning discharges. The vertical axis corresponds to altitude in the atmosphere, while the horizontal axis represents the horizontal distance from the lightning discharge. The CMC value created by each discharge is 140~C km, 350~C km and 560~C km. We show snapshots at 1~ms and 3~ms after the beginning of each discharge.
}
\end{figure}

\begin{figure}
\includegraphics[width=1\columnwidth]{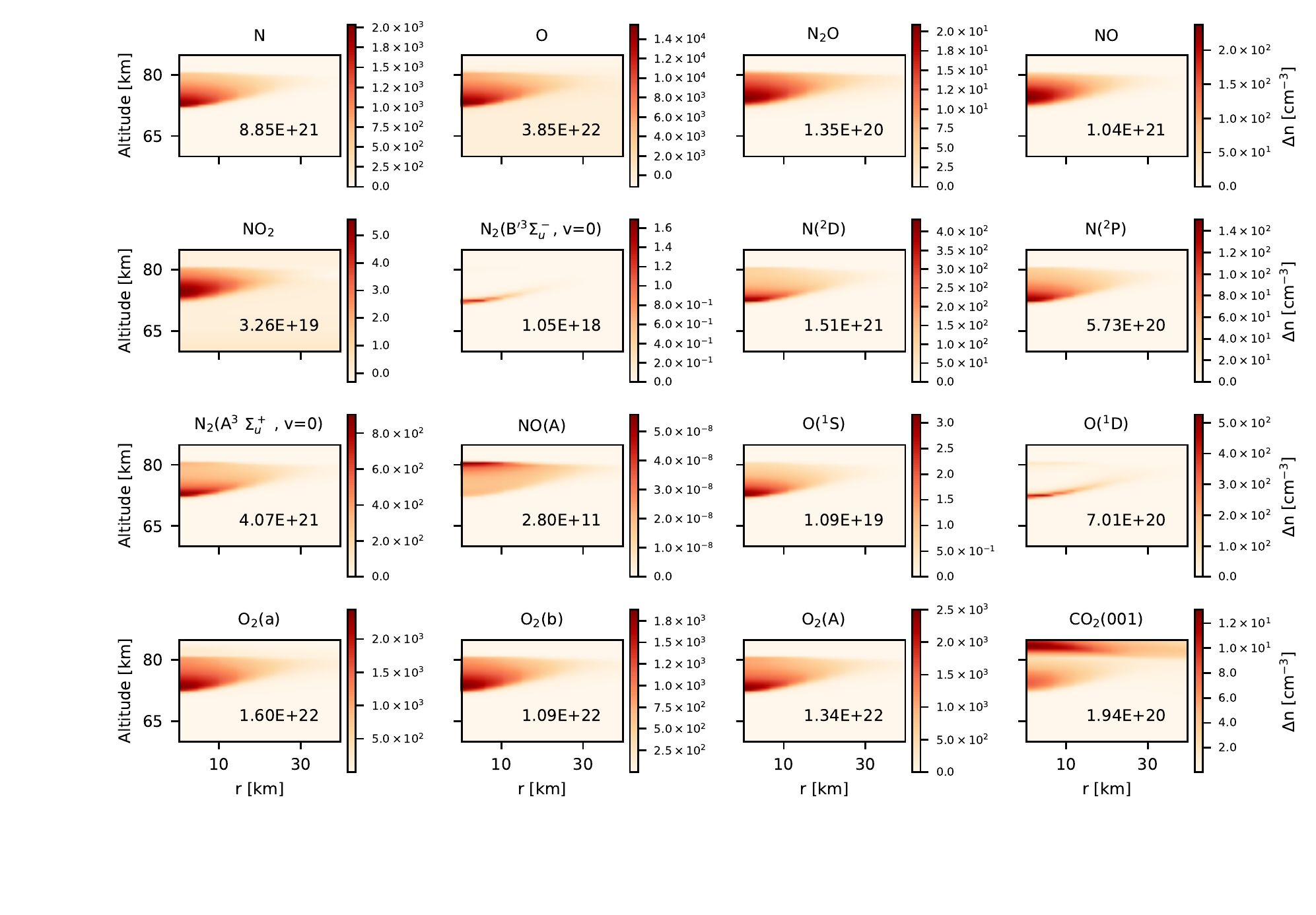}
\caption{\label{fig:species_2}
Variation (with respect to ambient values) of the density of some neutrals in the atmosphere of the Earth 3~ms after the beginning of a CG lightning discharge producing a CMC of 560~C km. The axes are the same as in Figure \ref{fig:Halos_Ered}. We show the total number of molecules created by the halo in the lower right corner of each subplot.
}
\end{figure}

\begin{figure}
\includegraphics[width=1\columnwidth]{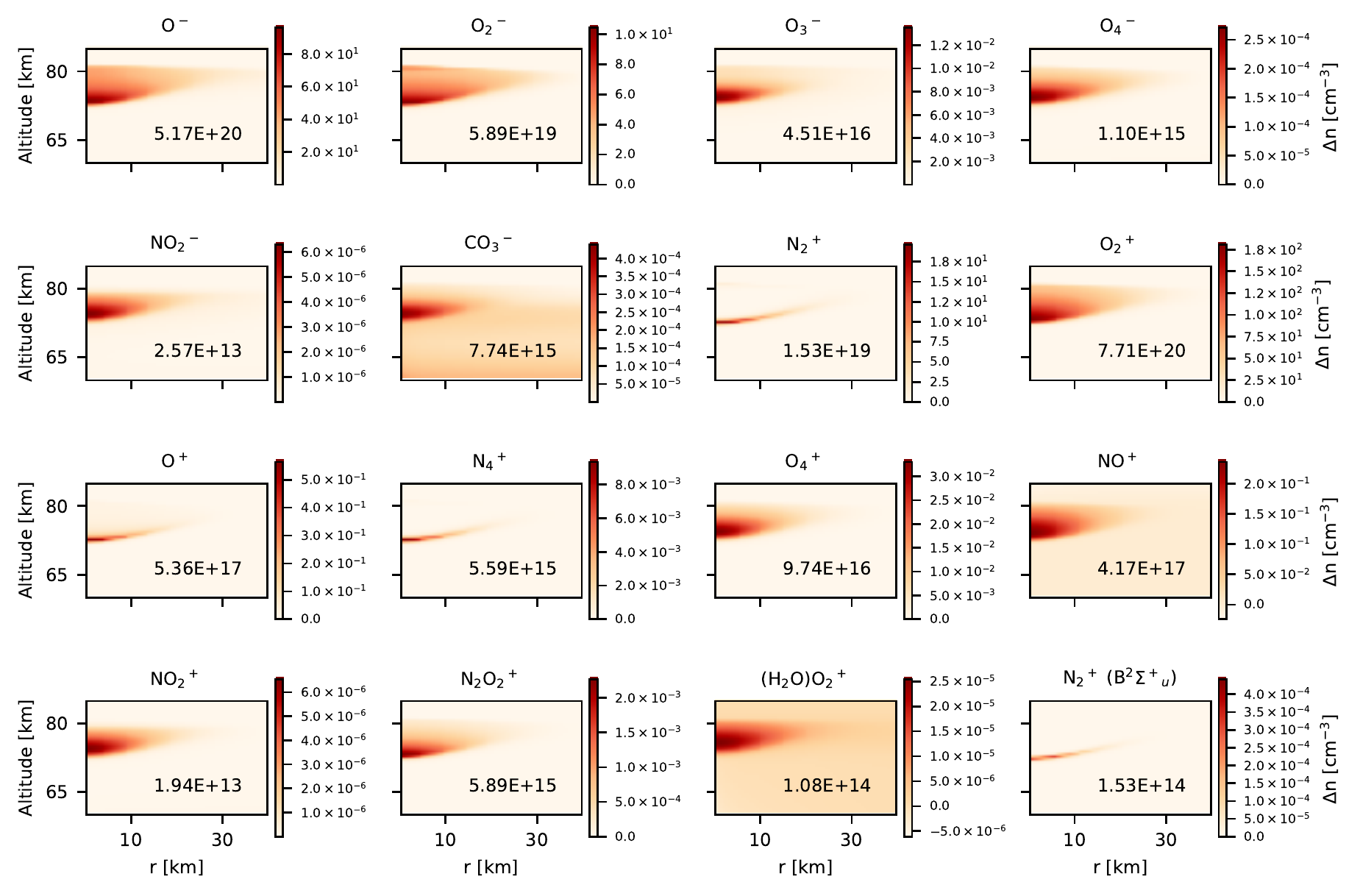}
\caption{\label{fig:species_1}
Density variation of some ions in the atmosphere of the Earth 3~ms after the onset of a CG lightning discharge producing a CMC of 560~C km. The axes are the same as in Figure \ref{fig:Halos_Ered}. We show the total number of molecules created by the halo in the lower right corner of each subplot. 
}
\end{figure}

Figure \ref{fig:Halos_Ered} shows the reduced electric field and the electron density in the upper atmosphere created by vertical CG lightning discharges with different CMCs. The reduced electric field reaches the breakdown value of $\sim$120~Td (or $\sim$120 $\times$ 10$^{-17}$ V cm$^2$) in the case of the two most energetic discharges, increasing the electron density in the lower ionosphere at an altitude of $\sim$80~km. Afterwards, this enhancement of electrons triggered by ionization contributes to screen the electric field, as can be clearly seen in the case of a lightning discharge with a CMC of 560~C km.

Let us analyze now the chemical influence of halos focusing on each species. Figures \ref{fig:species_2} and \ref{fig:species_1} summarize the density variation of some of the main neutrals and ions in the atmosphere of the Earth. It can be seen that the main chemical effect of halos is focused at altitudes around 75~km. Furthermore, the horizontal chemical influence extends up to 20~km from the center of the halo.

The density of some initial ground state neutrals suffers an enhancement in the center of the halo 3~ms after its onset. The atomic nitrogen N, whose background concentration  is negligible, increases by about 8.8 $\times$ 10$^{21}$ molecules. The increase of N is followed by other species like O, N$_2$O, NO$_2$, and NO, with increases with respect to their ambient values of $\sim$~0.7~$\%$, $\sim$~0.2~$\%$, $\sim$~0.1~$\%$ and $\sim$~0.01~$\%$, respectively.  The concentration of NO$_x$ would increase after the extinction of this halo, as the produced N atoms will rapidly be converted into NO and NO$_2$ after interacting with O$_2$.

The main processes that contribute to enhance the densities of N and O are the collisions of electrons with N$_2$ and O$_2$, respectively. The enhancement in the density of N$_2$O is due to the associative detachment of O$^-$ by N$_2$, while the increase of NO is influenced by processes that involve N($^{2}$D) and O$_2$. Finally, NO interacts with molecules containing O atoms to create NO$_2$.

\subsubsection{Model limitation: Possible sprite inception} \label{sec:limitation}

\begin{figure}
\includegraphics[width=1\columnwidth]{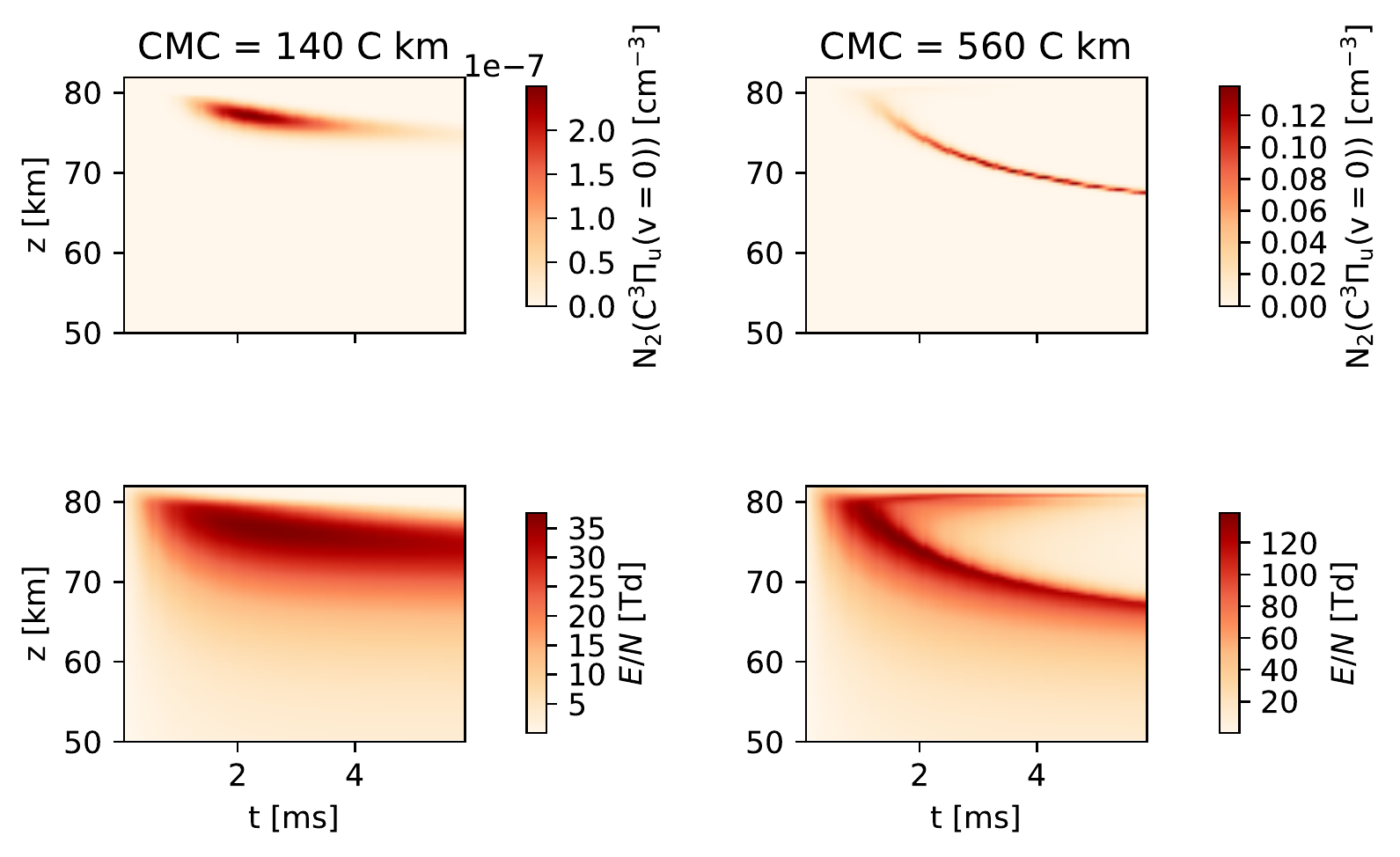}
\caption{\label{fig:emissions_t}
Temporal evolution of the density of the emitting species N$_2$(C$^{3}$$\Pi_u$(v = 0)) (top panels) and the reduced electric field (bottom panels) in a vertical column above the lightning discharge. Panels in the first and second column correspond to two different lightning discharges with CMCs of 140 C km and 560 C km, respectively.
}
\end{figure}

Figure~\ref{fig:emissions_t} shows the temporal evolution of the  N$_2$(C$^{3}$$\Pi_u$(v = 0)) density and the reduced electric field in a vertical column above two different discharges. For the weakest discharge, with a CMC of 140 C km, optical emissions due to N$_2$(C$^{3}$$\Pi_u$(v = 0)) disappear around 4 ms after the beginning of the discharge, while the reduced electric field is too low to produce emissions or ionization below 75 km of altitude. However, for higher CMCs the reduced electric field is above the breakdown value at altitudes below 75 km. In that region, the lack of electrons entails a dielectric relaxation time ($\tau_m = \epsilon_0  (e n_e \mu_e)^{-1}$) of tens of milliseconds, resulting in a long lasting halo that disagrees with observations  \citep{Marshall2006/JGRD, Kuo2013/JGR}. Probably, a sprite would appear in this situation, screening the electric field below 75 km of altitude. However, our model is not capable of describing the evolution of sprite streamers. 

%However, the situation is different when the lightning discharge is capable of producing greater electric fields. In the case of a discharge with a CMC of 560 C km, the region that contains the emissions becomes smaller than the spatial resolution for long times. In addition, the reduced electric field that produces the excitation of this species is not shielded at altitudes below 75 km, where the dielectric relaxation time is of the order of several milliseconds. According to our model, this situation would lead to an extremely thin halo with an unrealistic brightness duration of tens of milliseconds, as the induced current of electrons is not enough to shields the reduced electric field below 75 km of altitude. The description of this electrodynamical scheme would require a model capable of describing sprite inception.

\subsubsection{Long time halo simulation}
\label{longtime}
\begin{figure}
\includegraphics[width=1\columnwidth]{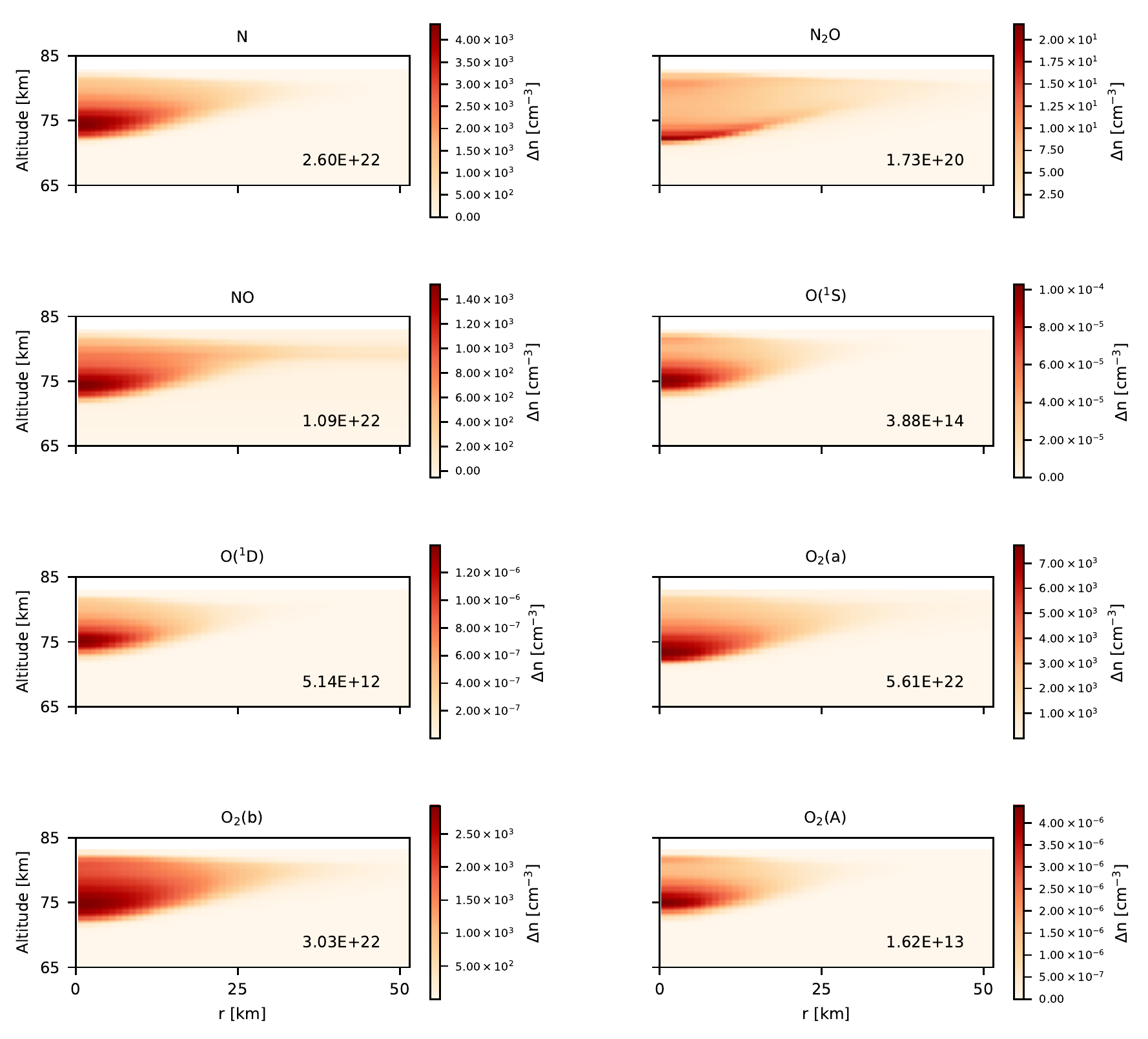}
\caption{\label{fig:species_3}
Variation of the density of some neutrals in the atmosphere of Earth 1 second after the beginning of a CG lightning discharge producing a CMC of 560~C~km. The axes are the same as in figure \ref{fig:Halos_Ered}. We show the total number of molecules created by the halo in the lower right corner of each subplot.
}
\end{figure}

\begin{figure}
\includegraphics[width=1\columnwidth]{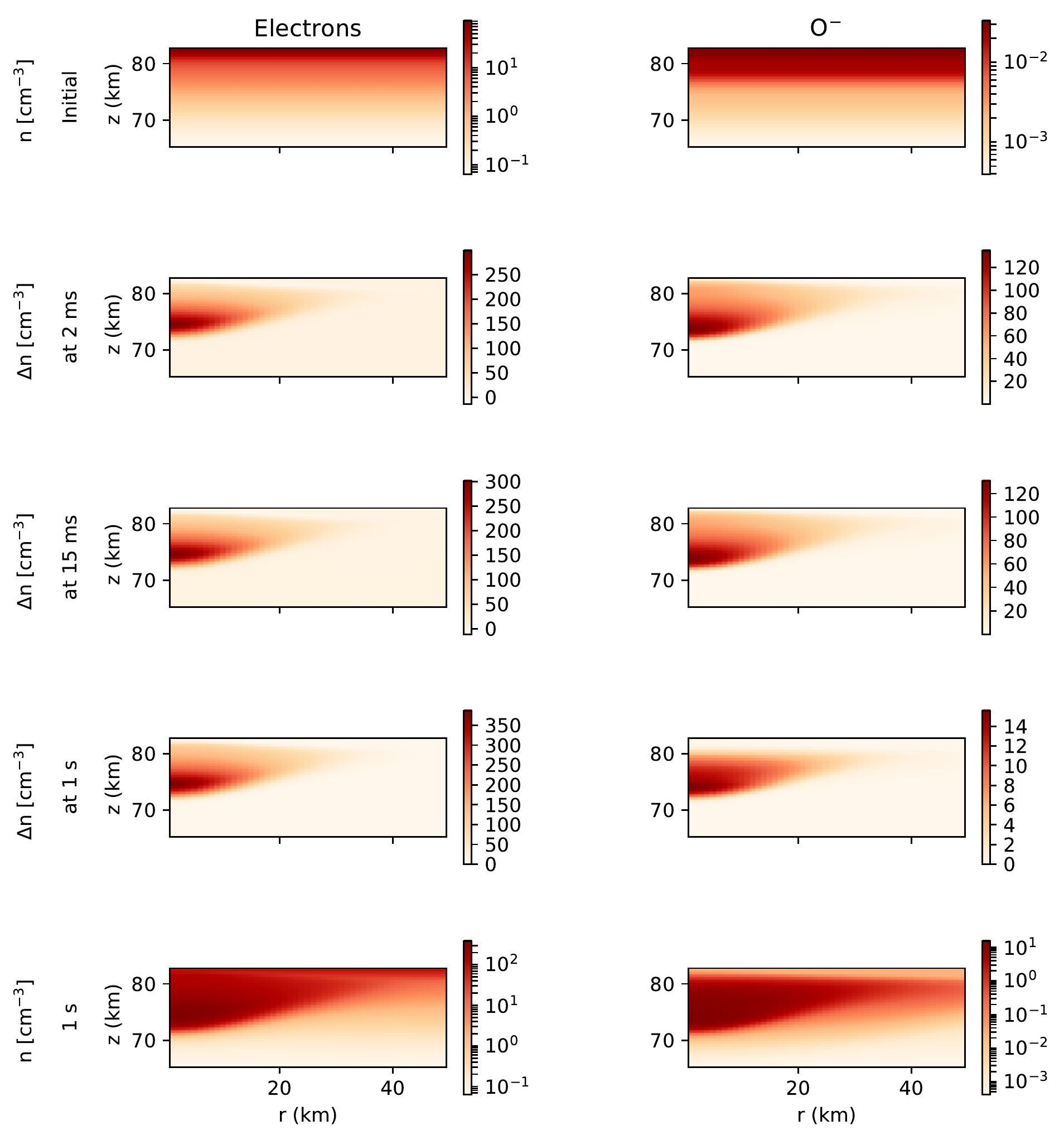}
\caption{\label{fig:species_4}
Evolution of the density of electrons and O$^{-}$ in the atmosphere of the Earth during 1~second after the beginning of a CG lightning discharge producing a CMC of 560~C~km. The axes are the same as in figure \ref{fig:Halos_Ered}. The first and the last rows shows the initial and final profiles, respectively. The second, third and fourth rows show the increase in the density at different times since the beginning of the lightning discharge.
}
\end{figure}

In section \ref{sec:limitation} we verified that the halo model fails for relatively long times ($\sim$6~ms) for lightning discharges producing a high CMC ($\sim$560~C~km), as it cannot simulate the development and spreading of sprite streamers. However, \cite{Kuo2013/JGR} reported some luminous halos without sprite inception triggered by lightning discharges with large CMCs, greater than 800~C~km. In this section we model halos produced by a lightning discharge that accumulates 80 C of charge during one millisecond, followed by another discharge that removes the accumulated charge in the next five milliseconds. This approach allows us to avoid sprite inception. In addition, the hypothesis whereby the charge on clouds is removed by a subsequent discharge would explain the exceptional single halos reported by \cite{Kuo2013/JGR}.

After the cloud charge removal, the only net charge present in the amosphere would be the sole halo induced charge in the mesosphere. According to our simulation, these charges would produce a maximum reduced electric field of 13~Td, quite lower than the breakdown field. We neglect the effect of this field and deactivate both the Poisson solver and the transport of charged particle in order to accelerate the calculations. In addition, we decrease the spatial resolution. This allows us to extend the simulation to predict the local chemical influence of lightning discharges in the lower ionosphere up to the scale of seconds. 

The most important variations in the density of neutrals 1~s after the halo onset are plotted in figure~\ref{fig:species_3}. Although the increase of ground neutrals is more than one order of magnitude lower than background densities shown in figure \ref{fig:equilibrium}, it is interesting to note and quantify the enhancements of some important species, such as N$_2$O and NO. The increase of these species densities with respect to background in the center of the halo 1~s after its onset is of $\sim$~0.2~$\%$ and $\sim$~0.1~$\%$, respectively. Despite the high relative enhancement of atomic nitrogen, the absolute increase is of the same order than NO, as can be seen in the number written in the first plot of figure~\ref{fig:species_3}. We can compare figures~\ref{fig:species_2} and~\ref{fig:species_3} to note that a long time simulation produces a larger ratio of NO$_x$ to N, as N is converted into NO$_x$.These results suggest that halos have a non-negligible local and regional chemical influence in the upper atmopshere near thunderstorms. \cite{Arnone2008/GeoRL} obtained a local enhancement of NO$_x$ produced by sprites streamers of 10~$\%$ at 52~km of altitude, increasing up to 60~km. Therefore, we can conclude that the chemical influence of a halo is between one and two orders of magnitude below the influence of a sprite.

We can also estimate the energy deposited in the mesosphere by a halo. Our models calculate the total flux of electrons produced by the lightning-generated electric field. Given both the temporal evolution of the flux of electrons and the electric field, the power deposited in the mesosphere can be calculated as the product of these two quantities and the total volume of the halo. Finally, the total deposited energy can be estimated knowing the duration of the event. This calculation leads us to estimate that the total amount of energy deposited in the mesosphere is about 10$^6$~J. Therefore, the production rate of NO by a halo can be approximated in terms of energy as 10$^{16}$ molecules of~NO/J, one order of magnitude lower than the production rate of NO by lightning, estimated in 10$^{17}$ molecules of~NO/J \citep{Price1997/JGR}.
The ISUAL instrument observations estimated an annual occurence of TLEs about 1.2$\times$10$^{7}$ \citep{Chern2014/AA}. The 6$\%$ of the observed TLEs by ISUAL were halos. The computed production of NO molecules by a halo together with the observation of halo occurrence by ISUAL allow us to estimate the total amount of NO created by halos in 2$\times$10$^{-7}$~teragrams of nitrogen per year (Tg~N~/~y). This value is quite below the estimated production of NO by lightning discharges, estimated between 5 and 9~Tg~N~/~y \citep{SchumannHuntrieser2007/SCP,Nault2017/JGR}. According to these number, the global chemical influence of halos is trivial. 

It is also worth analyzing the temporal evolution of electrons up to 1~second. Figure \ref{fig:species_4} shows the evolution of the density of electrons and O$^{-}$ in the atmosphere of the Earth. It can be seen how the O$^-$ is transformed into electrons between 15~ms and 1~s. The main chemical process that contributes to this transformation is the associative detachment reaction

\begin{linenomath*}
\begin{equation}
O^{-} + CO \rightarrow e + CO_2,
\end{equation}
\end{linenomath*}

that exceeds other associative detachment processes when the applied electric field is zero. However, at the very initial moment when the halo develops and the electric field is high, the rate of this reaction does not increase, since it does not depend on the electric field \citep{Moruzzi1968/TJCP}.

\subsubsection{Optical signature produced by halos}
\label{sec:opticalhalo}
\begin{figure}
\includegraphics[width=1\columnwidth]{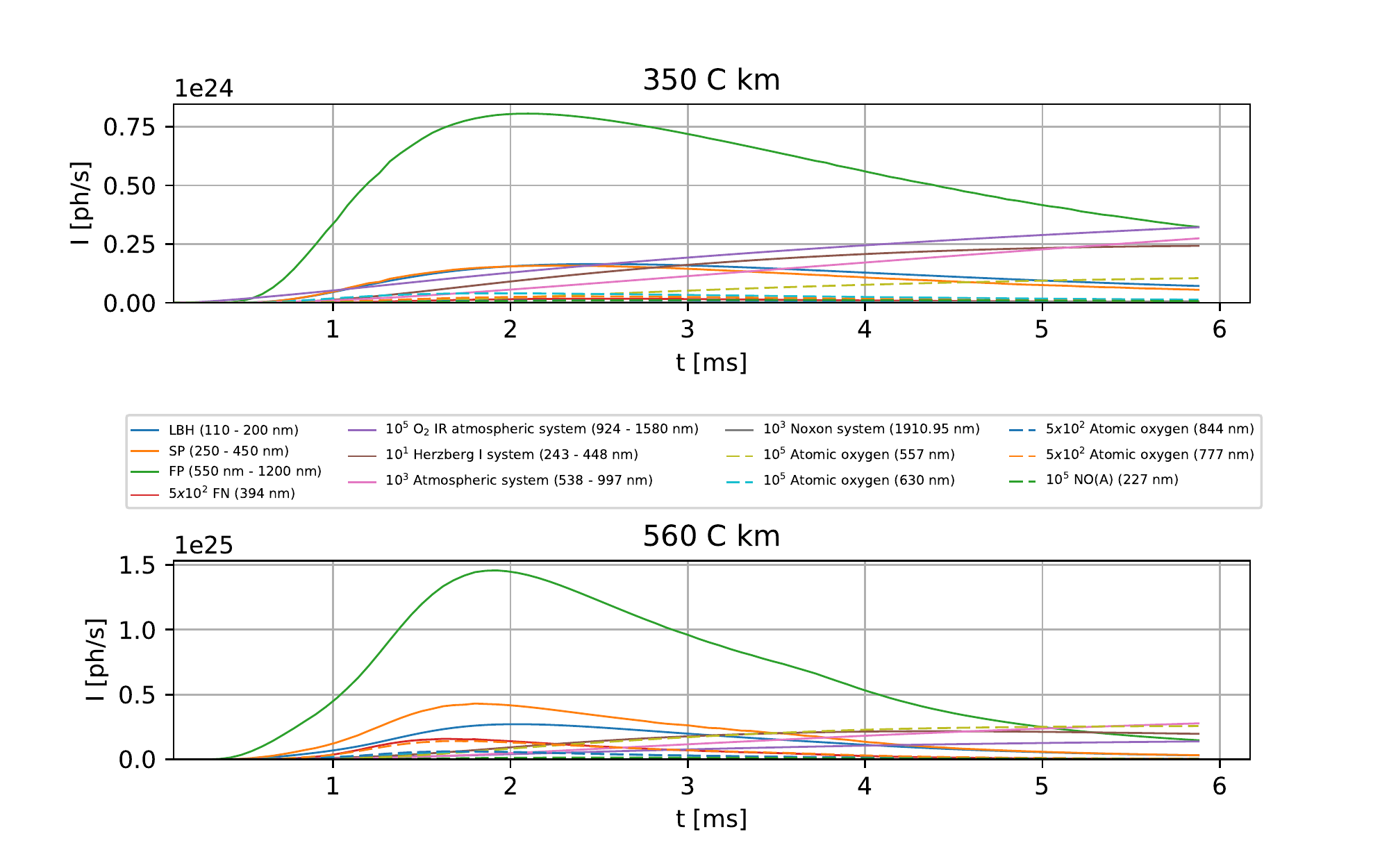}
\caption{\label{fig:halos_emissions}
Temporal evolution of the total emitted photons per second (for the main spectral bands) from halos. This figure shows results for two halos triggered by two CG lightning discharges producing total CMCs of, respectively, 350~C km and 560~C km. As can be seen in the legend box, some lines have been multiplied by different factors in an effort to plot all of them together. LBH, SP, FP and FN correspond to the Lyman-Birge-Hopfield band, second positive, first positive and first negative systems of molecular nitrogen, respectively.
}
\end{figure}

\begin{figure}
\includegraphics[width=1\columnwidth]{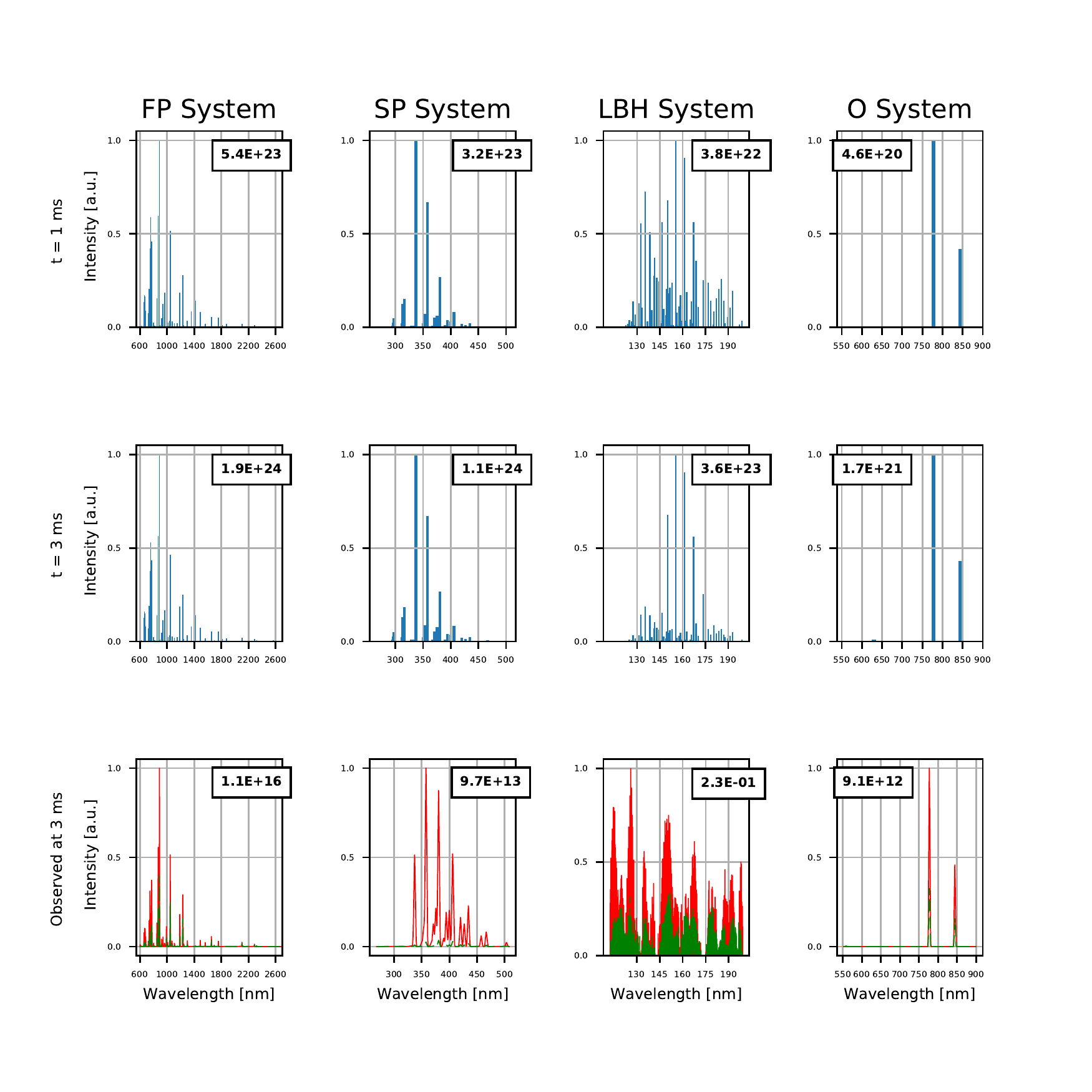}
\caption{\label{fig:halos_spectra}
Calculated spectra of halos for different spectral bands. The first and the second rows show different moments of the emission spectra at the source, while the third row shows the predicted observed spectra at 3~km (red solid line) and 275~m (green dashed line) over the sea and at a horizontal distance of 350~km (between the halo and the observer) 3~ms after the onset of the lightning discharge. We plot the intensity of the bands in arbitrary units, normalizing each subplot to the stronger transition in each band. The numbers in boxes correspond to photons per second in the case of emisson spectra, and photons per second and squared meters in the case of the predicted observed spectra.
}
\end{figure}

Figure~\ref{fig:halos_emissions} shows the temporal evolution of the main emission spectral bands in the first 6~milliseconds of halos produced by two different lightning discharges. In this plot, emissions of the first and second positive systems of molecular nitrogen, as well as the LBH band, have been obtained by summing the optical emission contribution from each vibrationally excited level. We analyze the optical emission from each species:

\begin{enumerate} 
\item Emissions from excited states of molecular nitrogen: 
The first positive system of the molecular nitrogen dominates over other bands, followed by the second positive system of the same molecule. The intensity of the LBH band is comparable to the intensity of the second positive system. The number of emitted photons per second in the first negative system are around 3~orders of magnitude lower than optical emissions of the first positive system.
It is interesting to note that the temporal position of each intensity peak is different as a consequence of the different lifetimes of each of the emitting species. 

\item Emissions from excited states of molecular oxygen: 
We obtain emitted photons from molecular oxygen in the spectral bands detailed in table~\ref{tab:species}. As can be seen in figure~\ref{fig:halos_emissions}, emissions from molecular oxygen are always between 1 and 6 orders of magnitude lower than emissions from the first positive system of molecular nitrogen.

\item Emissions from excited states of atomic oxygen and nitric oxide: 
figure~\ref{fig:halos_emissions} also shows the temporal emissions produced by radiative decay of electronically excited states of atomic oxygen (O) and nitric oxide (NO). In particular, our calculations indicate that some weak emissions corresponding to 227~nm, 557~nm, 630~nm, 777~nm and 844~nm would be produced by halos. However, these emissions would possibly be too weak to be detected by current instruments. The low background density of atomic oxygen and nitric oxide is the reason behind these weak optical emissions.
\end{enumerate}

Knowledge of the concentration of each vibrational level in N$_2$(B$^3$$\Pi _g$, v = 0,...,6), N$_2$(C$^3$$\Pi _u$, v = 0,...,4) and N$_2$(a$^1$$\Pi _g$, v = 0,...,15) allows us to build the Vibrational Distribution Function (VDF) of these electronically excited species \citep{Luque2011/JGRA}. Therefore, we can also derive systhetic emission spectrum of halos with vibrational resolution in the mentioned bands. Figure \ref{fig:halos_spectra} shows calculated spectra of halos corresponding to different spectral bands. Furthermore, we plot in this figure the predicted observed spectra at different observer altitudes (3~km and 275~m) and at a horizontal distance of 350~km from the halo. The software MODTRAN 5 \citep{berk2005modtran} has been used to calculate the optical transmittance of the atmopshere needed to derive the predicted observed spectra. In the case of a spacecraft observing from its orbit, the observed spectra would be similar to the emitted spectra at the source, given the low attenuation of light in the atmosphere at altitudes above 80~km of altitude.

Let us compare the obtained observed halo spectra with the ones previously calculated by \cite{Gordillo-Vazquez2011/JGR, Gordillo-Vazquez2012/JGR}. It can be seen how the first and second positive systems, as well as the LBH systems of the molecular nitrogen are in good agreement with halo spectra shown in \cite{Gordillo-Vazquez2011/JGR,Gordillo-Vazquez2012/JGR}.

\subsection{FDTD model results}

In this section we analyze elves triggered by CG lightning discharges, CIDs and EIPs. We use the models described in section~\ref{sec:FDTD} to investigate the local chemical impact and optical signature produced by EMPs in the lower ionosphere.

\subsubsection{Chemical impact and optical signature produced by CG lightning-generated elves}
\begin{figure}
\includegraphics[width=1\columnwidth]{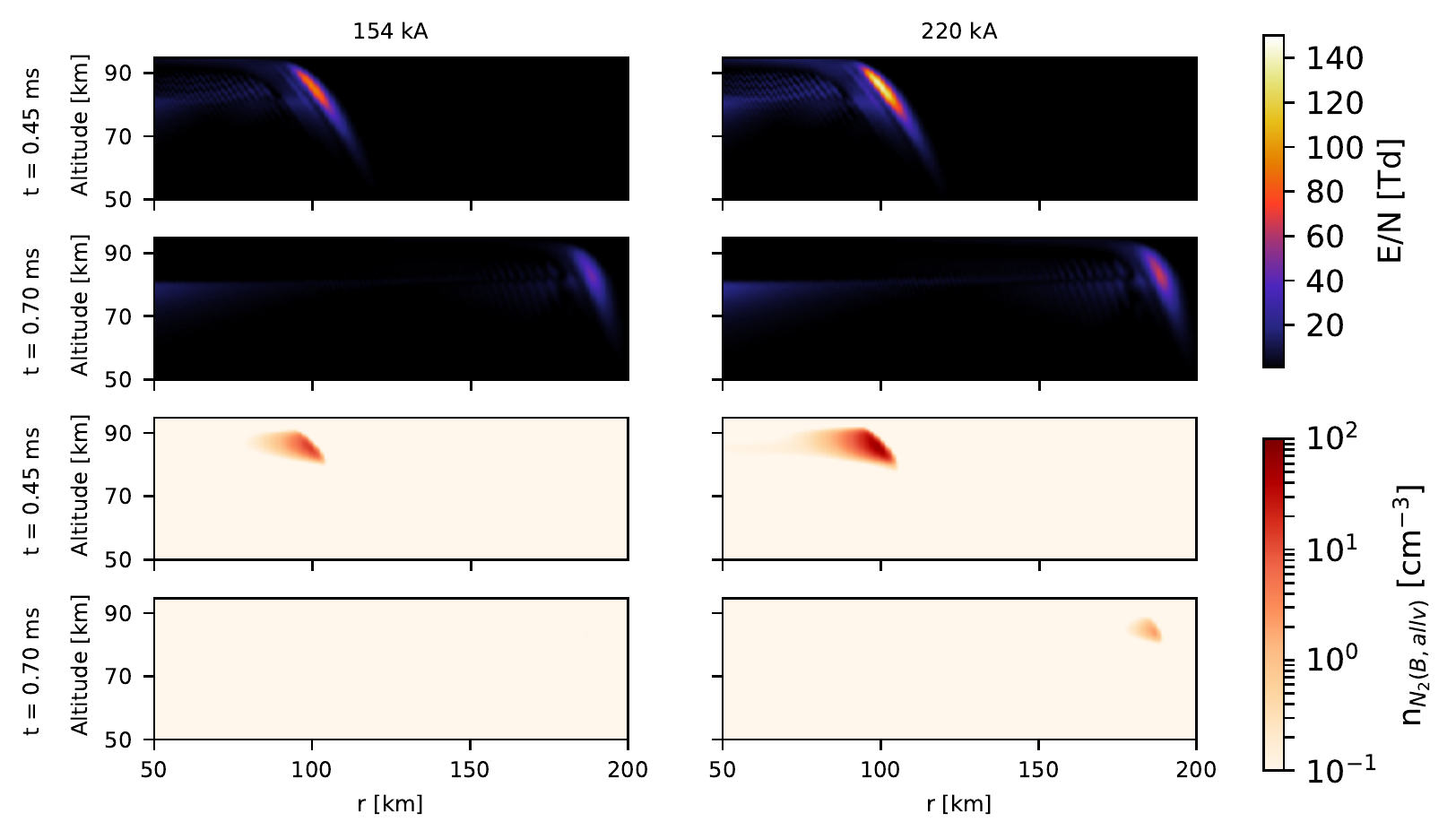}
\caption{\label{fig:Elves_Ered}
Reduced electric fields and densities of N$_2$(B$^3$$\Pi _g$, all v) in the upper atmosphere produced by two different vertical CG lightning discharges. The current peaks values associated to each discharge are 154~kA and~220~kA). We show results 0.45~ms and 0.70~ms after the beginning of each discharge.
}
\end{figure}

\begin{figure}
\includegraphics[width=1\columnwidth]{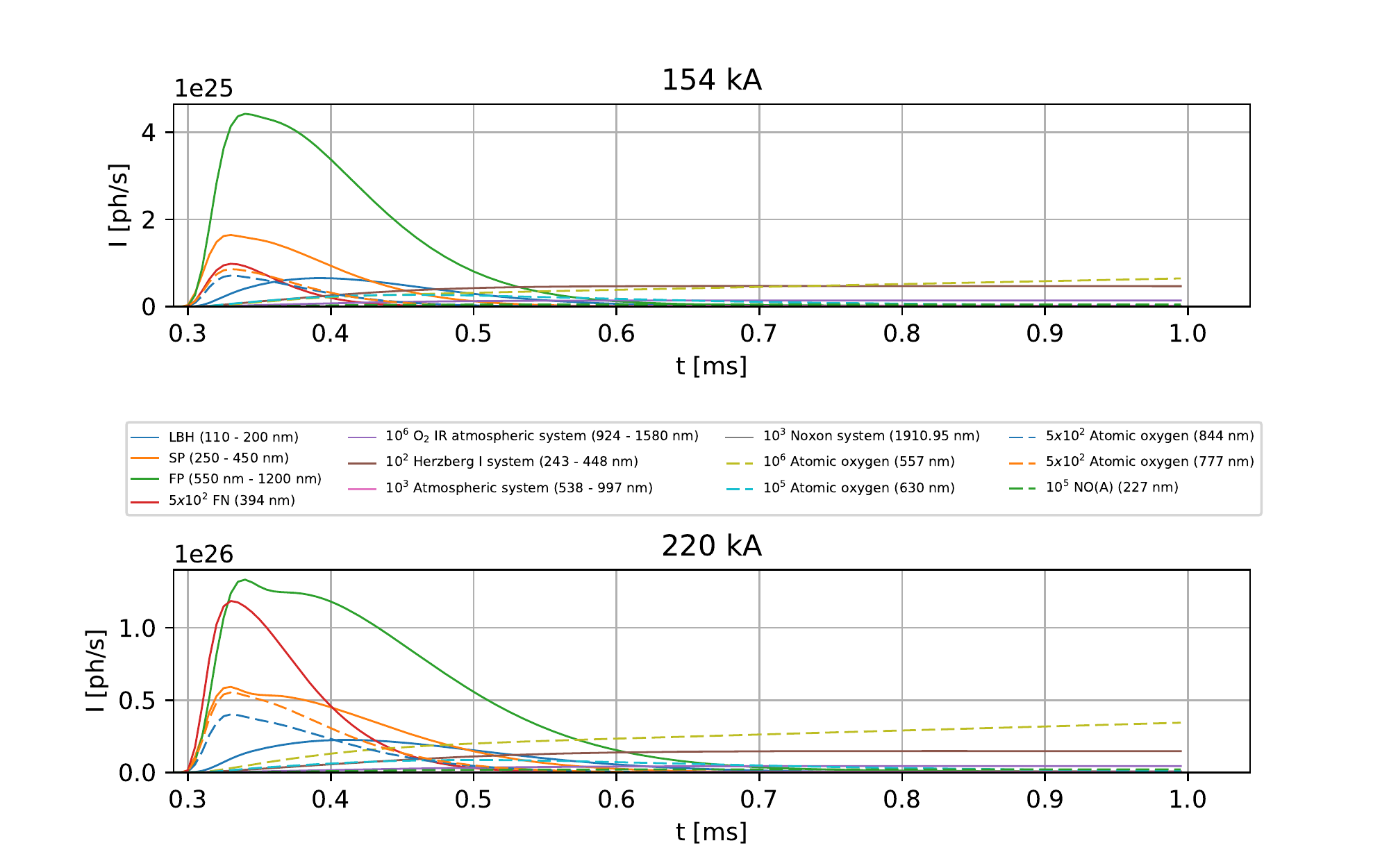}
\caption{\label{fig:elves_emissions}
Temporal evolution of the total emitted photons per second (for the main spectral bands) from elves. This figure shows results for two elves triggered by two CG lightning discharges producing current peaks of 154~kA and 220~kA. As in figure \ref{fig:halos_emissions}, some lines have been multiplied by different factors in an effort to plot all of them together. LBH, SP, FP and FN correspond to the Lyman-Birge-Hopfield band, second positive, first positive and first negative systems of the molecular nitrogen, respectively.
}
\end{figure}

\begin{figure}
\includegraphics[width=1\columnwidth]{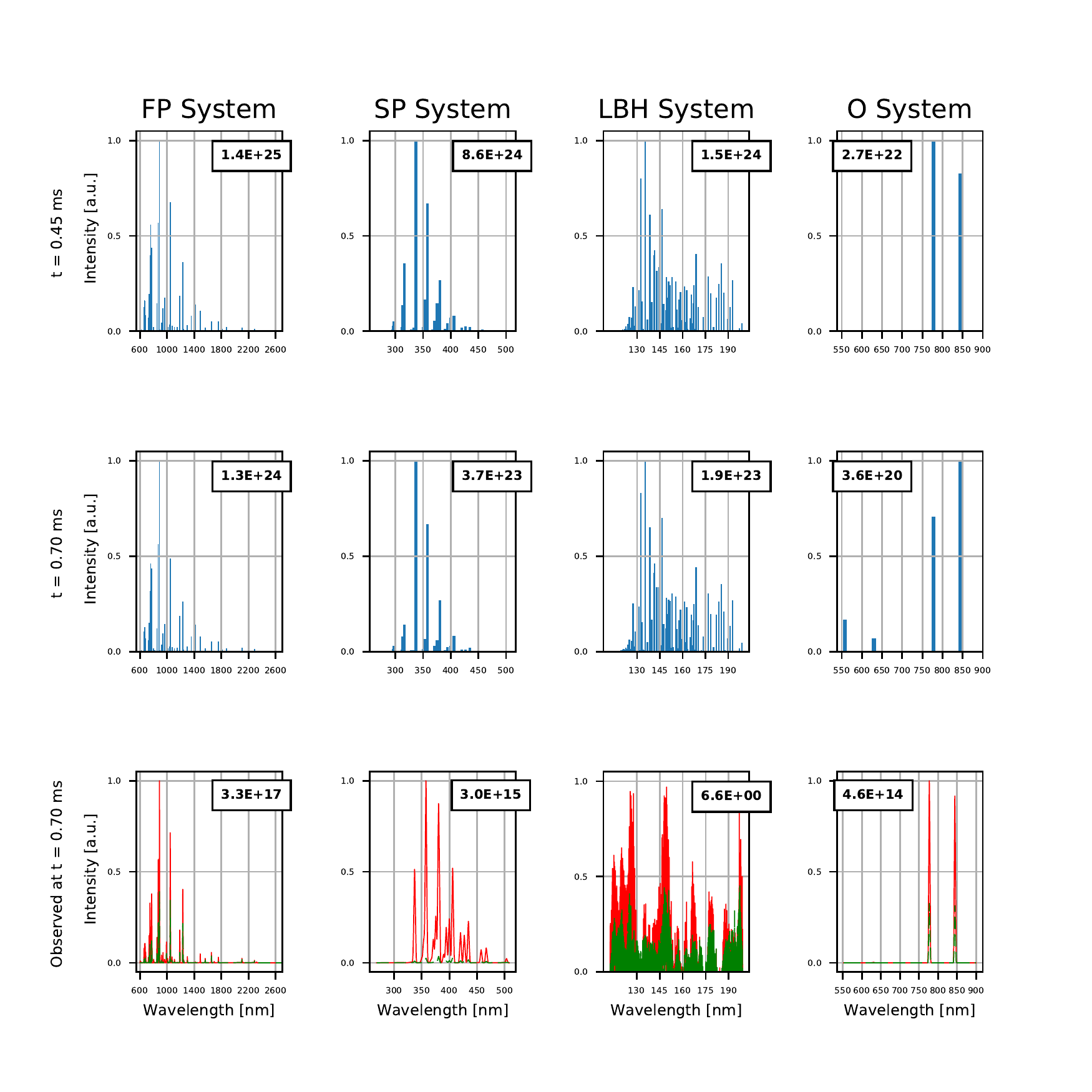}
\caption{\label{fig:elves_spectra}
Calculated spectra of elves produced by a CG lightning discharge with a current peak of 220~kA for different bands. The first and the second rows show the emission spectra at the source, while the third row shows the predicted observed spectra 0.70~ms after the onset of the lightning discharge at 3~km (red solid line) and 275~m (green dashed line) over the sea and at a horizontal distance of 350~km. We plot the intensity of the bands in arbitrary units, normalizing each subplot to the stronger transition in each band. The numbers in boxes correspond to photons per second in the case of emisson spectra, and photons per second and squared meters in the case of the predicted observed spectra.}
\end{figure}

Figure~\ref{fig:Elves_Ered} shows the reduced electric fields and densities of the emitting species N$_2$(B$^3$$\Pi _g$, all v) in the upper atmosphere produced by two different vertical CG lightning discharges. The pulse emitted by the temporal derivative of the current (\ref{current}) and the quasi-electrostatic field produced by the charge accumulation can be distinguished. The shape of the elves can be clearly distinguished in the last two rows of figure~\ref{fig:Elves_Ered} at around 88~km of altitude, where the density of N$_2$(B$^3$$\Pi _g$, all v) will produce toroidal-shaped optical emissions in the 337~nm spectral line by radiative decay. This figure also shows the quasi-electrostatic field produced by CG discharges in the first horizontal 50~km from the source, which would produce a halo.

In order to obtain the optical emissions produced by the elve itself without the influence of the halo, we calculate the part of the emissions produced from altitudes between 82~km to 90~km and from radial distances between 50~km and 200~km away from the source. The obtained temporal evolution of the main optical emissions triggered by the mentioned CG discharges are shown figure \ref{fig:elves_emissions}. The FP system of N$_2$ dominates over the rest of emissions, reaching its maximum 0.42~ms after the beginning of the discharge. The ratio between each emitted band is similar to the case of halos, except in the case of the first negative system of the most powerful discharge, whose relative importance increases given the high reduced electric field.

We also compute the emission spectra of elves. Figure~\ref{fig:elves_spectra} shows the main spectral bands where the elve can be detected. The observed spectra at points located 275~m and 3~km above the sea level and at a horizontal distance of 350~km from the elve is also shown in the figure.

Let us now analyze the local chemical impact of elves by estimating the production of NO molecules. We analyze the elves triggered by CG lightning discharges whose current peaks are 90~kA, 154~kA and 220~kA. The weakest of these discharges produces an elve slightly above the ISUAL detection threshold, estimated in parent lightning discharges with peak currents about 80~kA \citep{Kuo2007/JGRA, Chern2014/AA}, while the strongest discharge corresponds to a typical CG with a risetime of \SI{40}{\micro\second} \citep{Rakov2003/ligh.book}.  The simulated elves triggered by CG lightning discharges with current peaks of 90~kA, 110~kA and 220~kA create about 5$\times$10$^{17}$, 6$\times$10$^{19}$ NO and 4$\times$10$^{20}$ NO molecules, respectively, in agreement with \cite{Blaes2016/JGR}. We also compute (following the method described in section~\ref{longtime}) the total amount of energy locally deposited in the mesosphere by these three elves, obtaining 2$\times$10$^5$~J, 7$\times$10$^5$~J and 10$^6$~J, respectively. According to these quantities, the production rate of NO by elves in terms of energy would be in the range between 2.5$\times$10$^{12}$ molecules of~NO/J and 4$\times$10$^{14}$ molecules of~NO/J, that is, between 4 and 2~orders of magnitude below the NO production rate of halos. According to ISUAL observations, the global annual occurence of TLEs is about 1.2$\times$10$^{7}$ \citep{Chern2014/AA}, among which 74$\%$ are elves. The results of our elve simulations together with the observation of elves by ISUAL allow us to estimate that the total global amount of NO created by elves ranges between 10$^{-10}$~Tg~N~/~y and 10$^{-7}$~Tg~N~/~y. This quantity is between 10 and 7 orders of magnitude  lower than the estimated global annual production of NO by lightning discharges (between 5 and 9~Tg~N~/~y \citep{SchumannHuntrieser2007/SCP,Nault2017/JGR}). The global chemical influence of elves is then negligible.

%Regarding the local chemical impact of elves, we obtain a production of 5$\times$10$^{16}$ molecules by the elve triggered by a CG lightning discharge with a CMC of 1000~C~km. This production is almost 6 orders of magnitude lower than the NO molecules produced by halos, as the characteristic time of elves is shorter. We also compute (following the method described in section~\ref{longtime}) the total amount of energy deposited in the mesosphere by elves obtaining 1$\times$10$^6$~J. According to these quantities, the production rate of NO by an elve can be approximated in terms of energy as 10$^{14}$ molecules of~NO/J.

\subsubsection{Chemical impact and optical signature produced by CID-generated and EIP-generated elves}
\begin{figure}
\includegraphics[width=1\columnwidth]{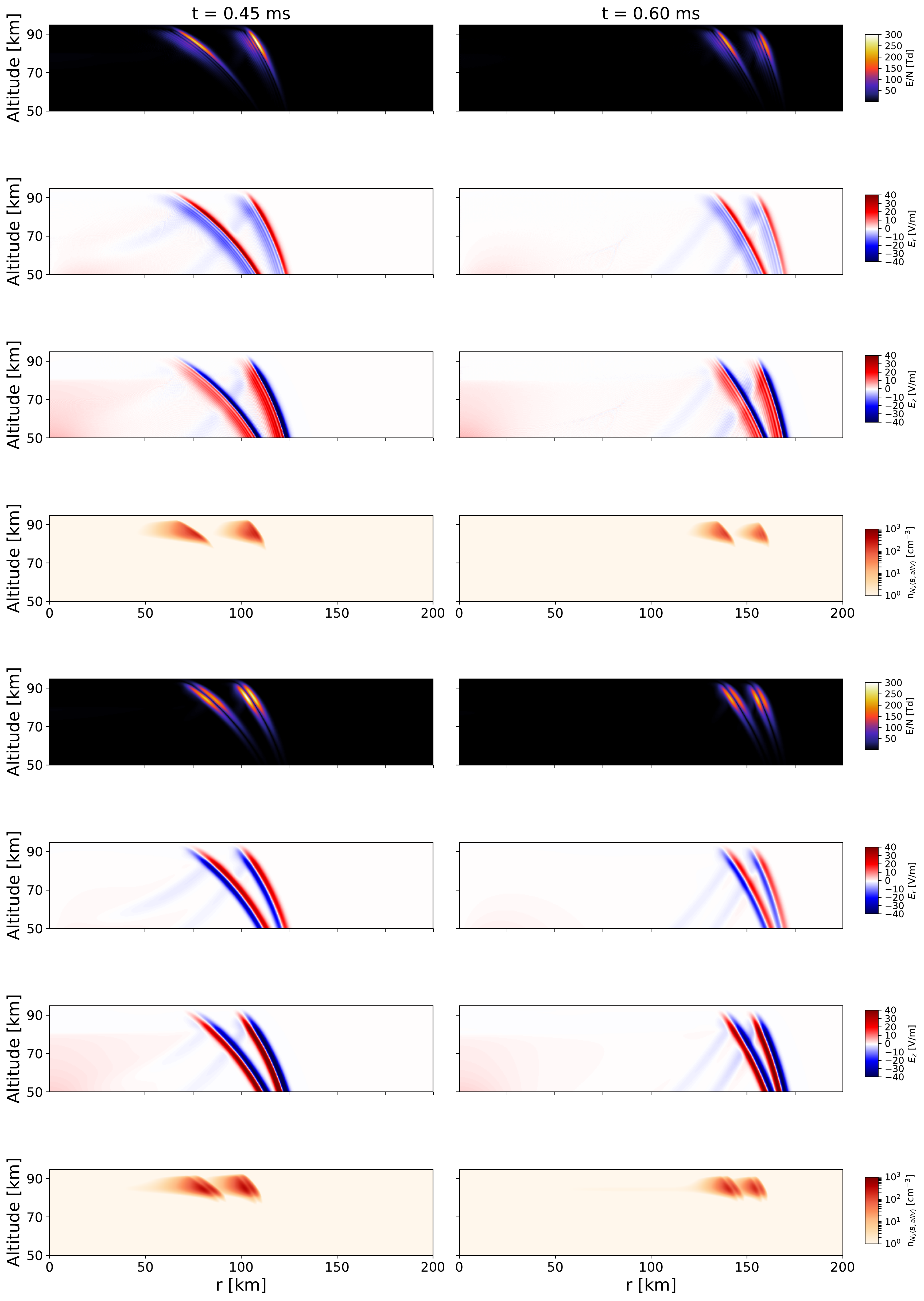}
\caption{\label{fig:CID_Ered}
Reduced electric field, $E_r$ and $E_z$ electric field components and the density of N$_2$(B$^3$$\Pi _g$, all v) in the upper atmosphere produced by CIDs and EIPs. We show results corresponding to 0.45~ms and 0.60~ms after the beginning of the discharge.
}
\end{figure}

\begin{figure}
\includegraphics[width=1\columnwidth]{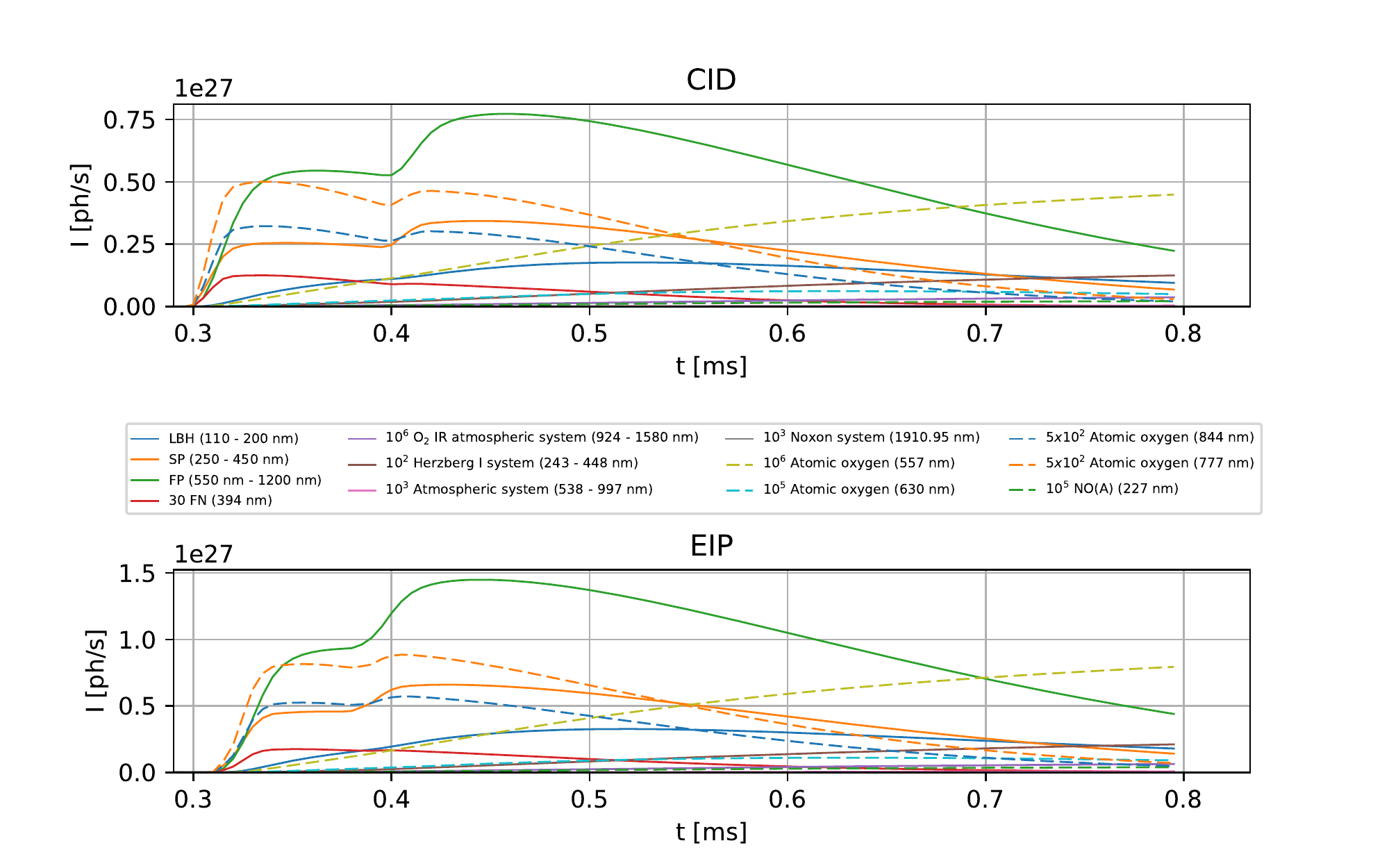}
\caption{\label{fig:CID_emissions}
Temporal evolution of the total emitted photons per second (for the main spectral bands) from elves produced by CIDs and EIPs. As in figure \ref{fig:halos_emissions}, some lines have been multiplied by different factors in an effort to plot all of them together. LBH, SP, FP and FN correspond to Lyman-Birge-Hopfield band, second positive, first positive and first negative systems of molecular nitrogen, respectively.
}
\end{figure}

\begin{figure}
\includegraphics[width=1\columnwidth]{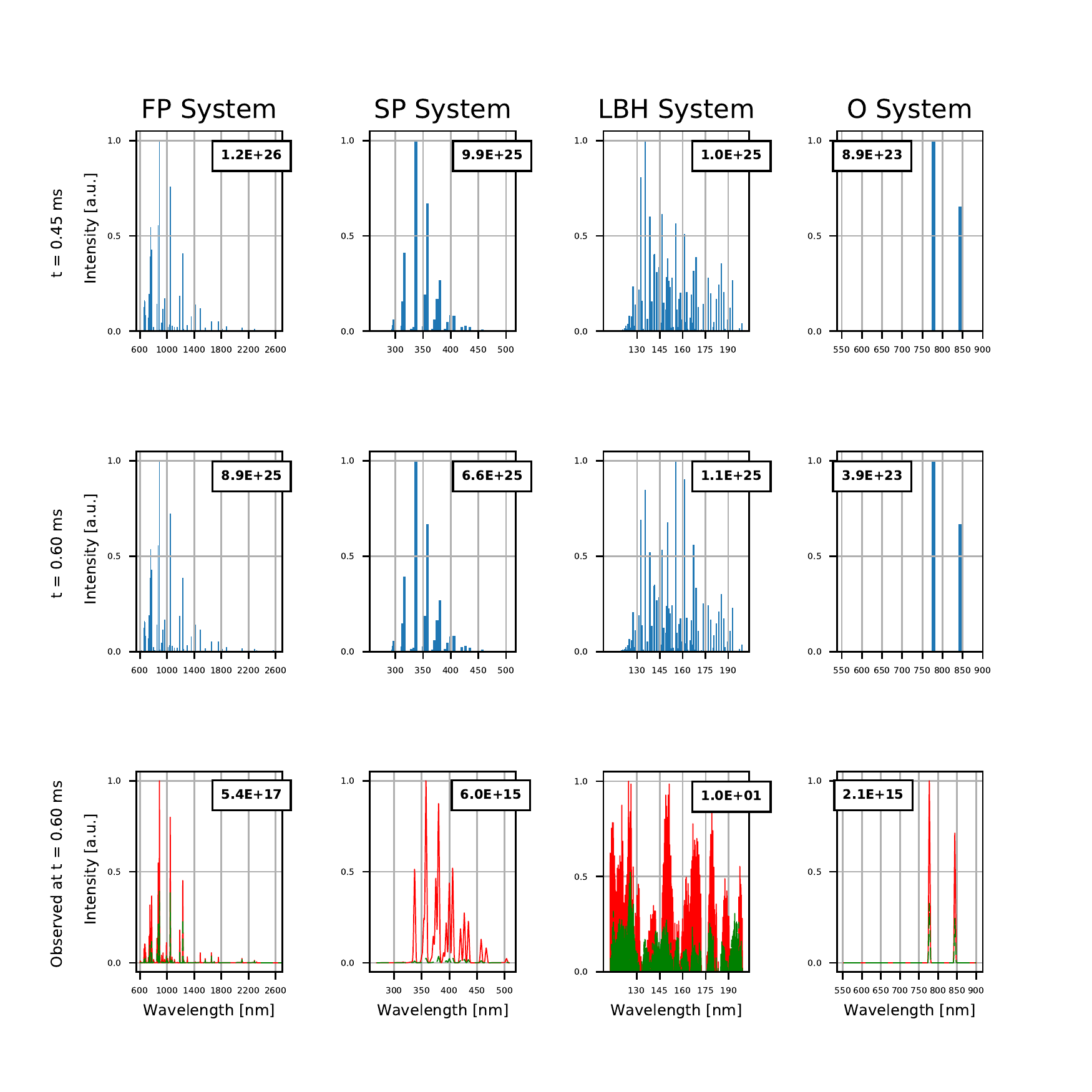}
\caption{\label{fig:CID_spectra}
Calculated spectra of elves produced by a CID. The first and the second rows show the emission spectra at the source, while the third row shows the observed spectra 0.6~ms after the onset of the CID at 275 m (red solid line) and 3 km (green dashed line) over the sea and at a horizontal distance of 350 km. We plot the intensity of the bands in arbitrary units, normalizing each subplot to the stronger transition in each band. The numbers in boxes correspond to photons per second in the case of emisson spectra, and photons per second and squared meters in the case of the predicted observed spectra.
}
\end{figure}

\begin{figure}
\includegraphics[width=1\columnwidth]{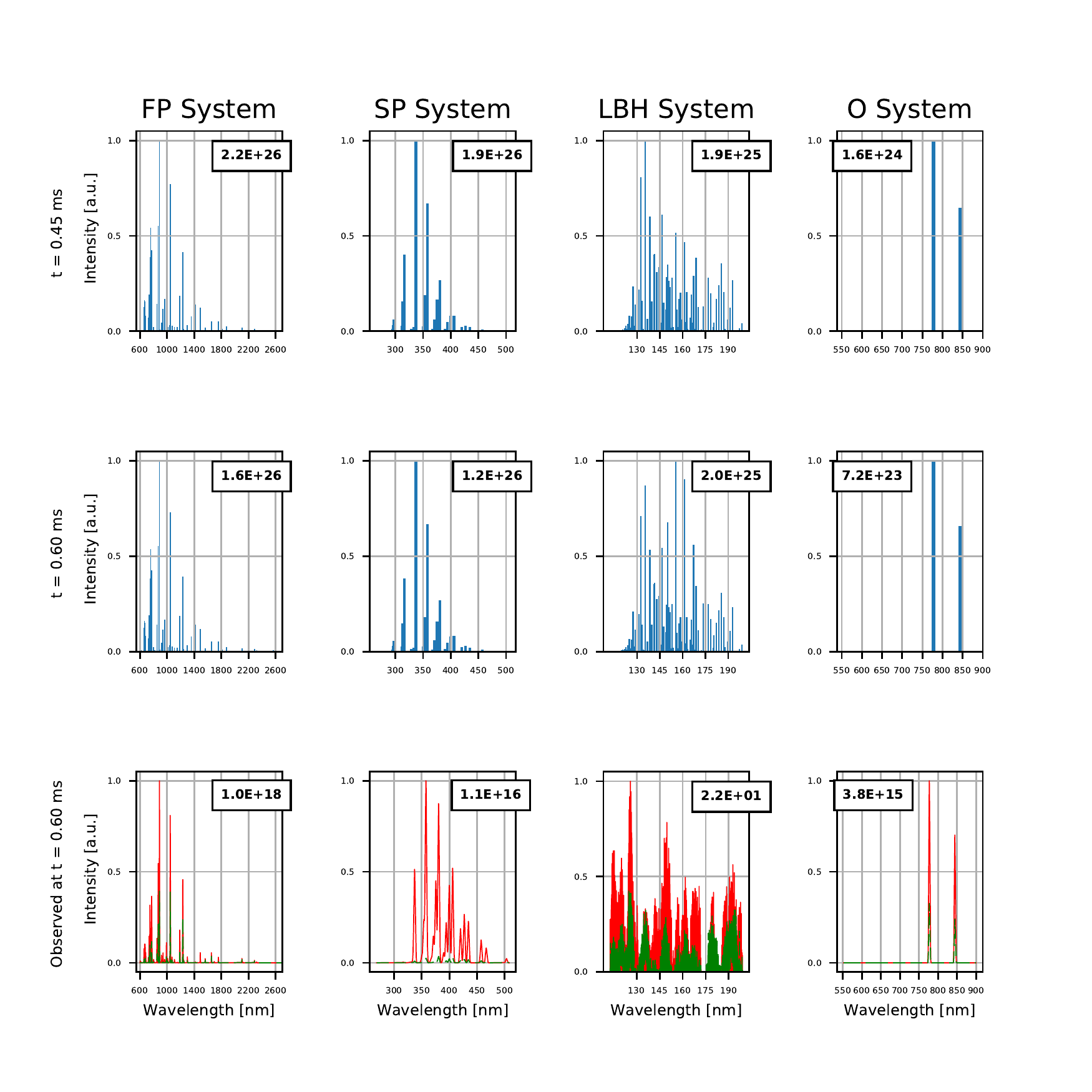}
\caption{\label{fig:EIP_spectra}
Calculated spectra of elves produced by an EIP. The first and the second rows show the emission spectra at the source, while the third row shows the observed spectra 0.6~ms after the onset of the EIP at 275 m (red solid line) and 3 km (green dashed line) over the sea and at a horizontal distance of 350 km. We plot the intensity of the bands in arbitrary units, normalizing each subplot to the stronger transition in each band. The numbers in boxes correspond to photons per second in the case of emisson spectra, and photons per second and squared meters in the case of the predicted observed spectra.
}
\end{figure}

As explained in section~\ref{sec:FDTD}, we investigate the local chemical impact and optical signature produced by CIDs and EIPs. The discharge current used as source is the one proposed by \cite{Watson2007/GRL} for the case of CIDs and by \cite{Liu2017/JGRAinpress} for the case of EIPs.

Figure~\ref{fig:CID_Ered} shows the reduced electric field and the density of N$_2$(B$^3$$\Pi _g$, all v) produced by both a CID and an EIP. This figure also shows the electric field components $E_r$ and $E_z$ produced by the discharges. As discovered by \cite{Newsome2010/JGR} and modelled by \cite{Marshall2015/GRL} and by \cite{Liu2017/JGRAinpress}, the EMP produced by CIDs and EIPs trigger a succession of two elves or elve ``doublet" as a consequence of the primary wave ground-reflexion. In addition, figure~\ref{fig:CID_Ered} shows how each pulse is formed by two sub-pulses with different polarization, due to the fast sign reversal of the derivative of the electric current at the source \citep{Watson2007/GRL,Liu2017/JGRAinpress}. The delay between the consecutive elves at a given distance from the center is different in the case of CIDs and EIPs as a consequence of the different altitudes of the current sources. The total number of NO molecules produced by these simulated CID and EIP-driven elves is about 10$^{21}$.

Figure~\ref{fig:CID_emissions} shows the obtained temporal evolution of the main optical emissions triggered by a CID and an EIP. As the current source temporal profile used in this work is different for the case of CIDs and EIPs \citep{Watson2007/GRL,Liu2017/JGRAinpress},  the temporal dependence of the optical emissions due to each event differs. The altitude of each event influences the delay between the peaks of maximum emissions, as can be seen after comparing the relative maximum of the emissions plotted in each planel of figure~\ref{fig:CID_emissions}.

We plot in figures~\ref{fig:CID_spectra} and~\ref{fig:EIP_spectra} the main spectral bands where the CID and the EIP-produced double elves emit photons. As in the last section, we predict the spectra as they would be observed at 275~m and 3~km above the sea level and at a horizontal distance of 350~km from the elve.

\section{Comparison of predicted spectra}

\begin{figure}
\includegraphics[width=1\columnwidth]{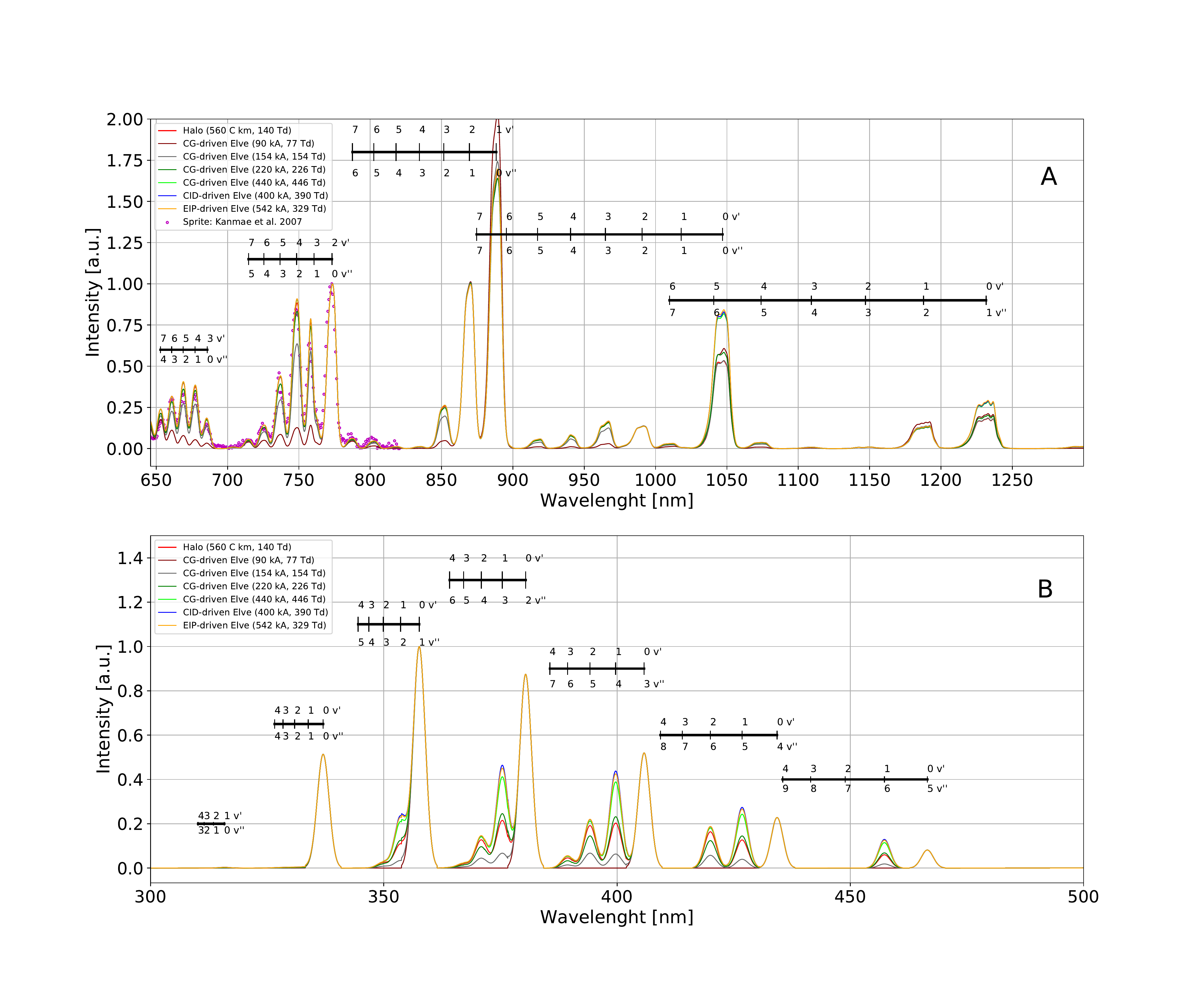}
\caption{\label{fig:spectra_comparison}
Calculated spectra of (A) the first positive system of N$_2$ and (B) the second positive system of N$_2$ for halos and elves produced by different discharges as seen by an observer located at an altitude of 3~km and a horizontal distance of 350~km. The magenta circles correspond to the sprite spectrum observed by \cite{Kanmae2007/GeoRL}. The observation of this sprite was performed from an altitude of 3.25~km and at horizontal distance of 350~km, and the observed region of the sprite was between 84~km and 86~km of altitude. The normalization of the spectra of the FP and the SP systems of N$_2$ spectra were done with respect to the (2,0) and the (0,1) transitions, respectively. We have selected the spectra at the moment of maximum emission of each TLE. The legend indicates the characteristics of the parent-lightning and the maximum reduced electric field reached inside each TLE.
}
\end{figure}

In this section we analyze and compare the predicted spectra of halos and elves (figures~\ref{fig:halos_spectra},\ref{fig:elves_spectra}, \ref{fig:CID_spectra} and~\ref{fig:EIP_spectra}). We also plot with more detail in figure~\ref{fig:spectra_comparison} the observed spectral bands corresponding to the FP and the SP systems of N$_2$ of each TLE, where we have added the spectra of a CG-driven elve produced by a lightning discharge with an extreme CMC of 1600~C~m (276~kA).

There are some slight differences between the spectra of halos, CG-driven elves, CID-driven elves and EIP-driven elves plotted in figures~\ref{fig:halos_spectra},\ref{fig:elves_spectra}, \ref{fig:CID_spectra} and~\ref{fig:EIP_spectra}. These differences can be attributed to the influence of the reduced electric field in the ionosphere. This is shown in figure~\ref{fig:spectra_comparison}, where the spectra of the CG-driven elves depend on the lightning CMC. Also the spectra of the elves generated by the 1600~C~km CG discharge, the CID and the EIP are similar.

%A slight difference in the emitted LBH band can be seen after comparing CG-produced halos and elves spectra (figures~\ref{fig:halos_spectra} and \ref{fig:elves_spectra}), as the dominating line changes from $\sim$155~nm in the case of halos to $\sim$135~nm in the case of elves. Each event has a different time scale and each emitting excited state also as a different decay constant, therefore the position of the dominant line depends on the characteristic time of each TLE. However, these differences cannot be appreciated in the calculated observed spectra.
%The slight differences between the spectra of CG-produced elves  (figures~\ref{fig:elves_spectra} and \ref{fig:CID_spectra}) are due to the different values of the reduced electric field reached in each case (see figures~\ref{fig:Elves_Ered} and~\ref{fig:CID_Ered}). Therefore, spectral differences between these two elves would depend on the source peak current rather than in the the kind of discharge that triggers them.

Observed spectra of halos are very noisy as a consequence of their low luminosity, as the one recorded shown in \cite{Wescott2001/JGR/1} and later on analysed by \cite{Gordillo-Vazquez2011/JGR}. However, we can compare our results with the optical sprite spectrum observed by \cite{Kanmae2007/GeoRL}. The sprite region observed by \cite{Kanmae2007/GeoRL} was located at an altitude between 84~km and 86~km, while the observation point was located at a mountain with an altitude of about 3~km and at a horizontal distance of 350~km from the TLE. Figure~\ref{fig:spectra_comparison} shows a comparison between the predicted spectra for the first and second positive systems of N$_2$ of halos and elves together with the sprite-spectrum observed by \cite{Kanmae2007/GeoRL}. It can be seen that the predicted spectra of the FP system of N$_2$ agrees with reported observations.

\section{Conclusions}
\label{conclusions}

Tropospheric electrical discharges such as CG lightning, CIDs and EIPs can produce mesospheric optical emissions known as TLEs. The electromagnetic fields that produce TLEs can also  trigger a cascade of chemical reactions producing a local chemical impact in the mesosphere. The main aim of this work has been to contribute to the knowledge of the characteristics of halos and elves, some of the most frequent TLEs, as well as to quantify their chemical signature. To achieve our goal, we have developed two different self-consistent models based on previus works \citep{Inan1991/GRL, Pasko1995/GeoRL} to study TLEs triggered by a single lightning discharge. Both models calculate the temporal evolution of more than 130~species in the lower ionosphere interacting through over 1000~chemical reactions, some of which are triggered by the lightning-produced electric field in the upper mesosphere and lower ionosphere. We have calculated the vibrational distribution of some electronically excited states of N$_2$ to obtain a detailed description of the FP and SP systems of the N$_2$ and the N$_2$ LBH bands to calculate synthetic spectra. In addition, we have considered the rotational structure of the FP system of N$_2$ and its corresponding spectrum. Finally, we have computed the effect of air absorption in the emitted optical emissions, predicting the observed spectra of the simulated TLEs at different distances from the source. This approach has enabled us to compare the characteristics of halos and elves produced by different tropospheric discharges.

The first developed model has allowed us to predict the spectra of a single halo and its chemical impact in the mesosphere after one second of being triggered. To perform this long-time simulation of a single halo, we have avoided the sprite inception problem simulating a second discharge that removes the electric field produced by the first one several milliseconds later.
The calculated spectra agree with previous model results \citep{Gordillo-Vazquez2011/JGR,Gordillo-Vazquez2012/JGR} and with the spectra of the FP system of N$_2$ of sprites detected by \cite{Kanmae2007/GeoRL} and more recently with the high resolution sprite spectra reported by \cite{Gordillo-Vazquez2018/JGR}. In addition, our models predict a non-negligible enhancement of local mesospheric N$_2$O, NO and metastable species as a consequence of this glow discharge. Future observations are needed to confirm the local chemical impact of halos and to establish their possible regional or global influence in the mesosphere. We estimate a global production of NO due to halos and elves of the order of 10$^{-7}$~Tg~N/y, which is not significant for global scale chemistry.

We have developed another model to simulate elves produced by CG lightning discharges, CIDs and EIPs. We have estimated for the first time the optical spectra of elves triggered by CIDs and EIPs. According to our results, it is not the type of discharge what influences the observed spectra of the produced TLEs but the value of the reduced electric field reached in the lower ionosphere. 
Despite the similarities in the spectra of elves produced by CG lightning discharges and CIDs or EIPs, the appearance of the elves produced by each type of discharge would be different. The former would usually be detected as single elves, while the latter would appear as double elves, as previosly investigated by \cite{Marshall2015/GRL} and \cite{Liu2017/JGRAinpress}.

\section*{Acknowledgement}

This work was supported by the Spanish Ministry of Science and Innovation, MINECO under projects ESP2015-69909-C5-2-R  and ESP2017-86263-C4-4-R and by the EU through the H2020 Science and Innovation with Thunderstorms (SAINT) project (Ref.~722337) and the FEDER program. FJPI acknowledges a PhD research contract, code BES-2014-069567.  AL was supported by the European Research Council (ERC) under the European Union’s H2020 programme/ERC grant agreement 681257.  The simulation data and plot codes presented here are available from figshare repository at https://figshare.com/s/f1c9f6c7bc728d6669dd. Alternatively, requests for the data and codes used to generate or displayed in figures, graphs, plots, or tables are also available after a request is made to the authors F.J.P.I (fjpi@iaa.es), A.L (aluque@iaa.es), or F.J.G.V (vazquez@iaa.es).

\newcommand{\pra}{Phys. Rev. A} 
\newcommand{\jgr}{J. Geoph. Res. } 
\newcommand{\jcp}{J. Chem. Phys. } 
\newcommand{\ssr}{Space Sci. Rev.} 
\newcommand{\planss}{Plan. Spac. Sci.} 
\newcommand{\pre}{Phys. Rev. E} 
\newcommand{\nat}{Nature} 
\newcommand{\icarus}{Icarus} 
\newcommand{\ndash}{-}

\clearpage

\tiny

\begin{longtable}{|c|c|c|c|}
\hline  \multicolumn{1}{|c}{\textbf{Neutrals}} & \multicolumn{1}{|c|}{\textbf{Negative charged particles}} & \multicolumn{1}{c|}{\textbf{Positive charged particles}} & \multicolumn{1}{c|}{\textbf{Emission bands}}  \\

\endfirsthead
\multicolumn{1}{|c}{\textbf{Neutrals}} & \multicolumn{1}{|c|}{\textbf{Negative charged particles}} & \multicolumn{1}{c|}{\textbf{Positive charged particles}} & \multicolumn{1}{c|}{\textbf{Emission bands}} \\
\hline
\endhead
\caption{Species considered in the chemical scheme and emission bands.} \label{tab:species} \\

\endlastfoot

\hline
N$_2$(X$^1$ $\Sigma _g ^+$, v = 0,...,10),  N$_2$(A$^3$ $\Sigma _u ^+$ , v = 0,...,16),  N$_2$(B$^3$ $\Pi _g$ , v = 0,...,6),  & e, O$^-$,  & N$^+$, N$_2^+$,  N$_2^+$(B$^2\Sigma_u^+$), N$_3^+$,    & FP and SP system of N$_2$: (550 nm - 1200 nm) and  (250 nm - 450 nm) \\

N$_2$(C$^3$ $\Pi _u$ , v = 0,...,4), N$_2$(a$^1$ $\Pi _g$ , v = 0,...,15),   N$_2$(W$^3$ $\Delta _u$, v = 0,...,3), &  NO$^-$, NO$_2^-$,   &  N$_4^+$, O$^+$, O$_2^+$, O$_4^+$,   NO$^+$,  & FNS of N$_2$$^+$ (391.4 nm) \\

 N$_2$(E$^3$ $\Sigma ^+ _g$), N$_2$(w$^1$ $\Delta _u$),  N$_2$(a$^{\prime 1}\Sigma^-_u$),  N$_2$(a$^{\prime\prime 1}\Sigma^+_g$), N$_2$(B$^{\prime 3}\Sigma_u^-$, v = 0, 1) & CO$_3^-$, CO$_4^-$ & NO$_2^+$,  N$_2$O$^+$,  N$_2$O$_2^+$,  & LBH band of N$_2$ (110 nm - 200 nm) \\

NO(A$^2\Sigma^+$), N, N($^2$D), N($^2$P), O$_2$, O$_2$(A$^3\Pi_u^+$),  O$_2$(b$^1\Pi_g^+$),  & O$_2^-$, O$_3^-$ ,  & N$_2$NO$^+$,  O$_2$NO$^+$, & Emissions from molecular oxygen (538 nm-1580 nm, 1908 nm, 243 nm-488 nm) \\

 O$_2$(a$^1\Delta_g$), O, O($^1$S), O($^1$D), O($^3$P), O($^5$P), O$_3$, CO, &   NO$_3^-$  & (H$_2$O)O$_2^+$, (H$_2$O)H$^+$, & Noxon system (1908 nm)  \\

NO, NO$_2$, NO$_3$, N$_2$O, N$_2$O$_5$, H, OH, H$^*$, H$_2$O, HO$_2$,  & O$_4^-$  &  (H$_2$O)$_2$H$^+$, (H$_2$O)$_3$H$^+$,  & Herzberg I (243 nm - 488 nm) system \\

H$_2$O$_2$, HNO$_3$, CO$_2$, CO$_2$(001), CO$_2$(100),  CO$_2$(010),  &    &  (H$_2$O)$_4$H$^+$, (H$_2$O)OHH$^+$, & Emissions from atomin oxygen (557 nm), (630 nm), (777 nm) and (844 nm)  \\

CO$_2$(02$^0$0), CO$_2$(02$^2$0), CO$_2$(03$^1$0), CO$_2$(03$^3$0), CO$_2$(11$^1$0)  &  &  &  \\

\hline

\end{longtable}
\normalsize

\end{document}